\documentclass[a4paper,fleqn,useAMS,usenatbib]{mnras}

\usepackage{newtxtext,newtxmath}

\usepackage[T1]{fontenc}
\usepackage{ae,aecompl}

\usepackage{times}

\usepackage{graphicx}	
\usepackage{amsmath}	
\usepackage {threeparttable} 

\newcommand{\cms}{\,cm$^{-2}$}

\newcommand{\mbh}{M$_{\mathrm{BH}}$\,}
\newcommand{\msun}{M$_{\sun}$\,}

\newcommand{\nh}{$N_{\mathrm{H}}$\,}
\newcommand{\lam}{$\lambda_{\mathrm{Edd}}$\,}
\newcommand{\nhl}{$N_{\mathrm{H}} - \lambda_{\mathrm{Edd}}$\,}
\newcommand{\ar}{$a_{\mathrm{rad}}(r)$\,}
\newcommand{\arm}{$\langle a_{\mathrm{rad}}\rangle$\,}
\newcommand{\agm}{$\langle a_{\mathrm{grav}}\rangle$\,}
\newcommand{\anh}{$\langle a_{\mathrm{rad}}\rangle - N_{\mathrm{H}}$\,}
\newcommand{\lbol}{$L_{\mathrm{bol}}$\,}
\newcommand{\led}{$L_{\mathrm{Edd}}$\,}
\newcommand{\rin}{$R_{\mathrm{in}}$\,}
\newcommand{\rsh}{$R_{\mathrm{sh}}$\,}
\newcommand{\com}{$\sigma_{\mathrm{T}}$\,}
\newcommand{\tmax}{$T_{\mathrm{max}}$\,}
\newcommand{\risco}{$r_{\mathrm{ISCO}}$\,}
\newcommand{\rs}{$r_{\mathrm{s}}$\,}

\title[Radiative feedback from dusty AGN]{Radiation pressure-driven outflows from dusty AGN}

\author[N. Arakawa et al.]{
N. Arakawa,$^{1, 2}$\thanks{E-mail: na489@cam.ac.uk}
A. C. Fabian,$^{1}$
G. J. Ferland$^{3}$
and W. Ishibashi$^{4}$
\\
$^{1}$Institute of Astronomy, University of Cambridge, Madingley Road, Cambridge CB3 0HA, UK\\
$^{2}$Kavli Institute for Cosmology, University of Cambridge, Madingley Road, Cambridge CB3 0HA, UK\\
$^{3}$Department of Physics, University of Kentucky, Lexington KY 40506, USA\\
$^{4}$Physik-Institut, Universitat Zurich, Winterthurerstrasse 190, 8057 Zurich, Switzerland
}

\date{Accepted 2022 October 17. Received 2022 October 17; in original form 2022 August 25}

\pubyear{2022}

\begin{document}
\label{firstpage}
\pagerange{\pageref{firstpage}--\pageref{lastpage}}
\maketitle

\begin{abstract}
	Radiation pressure-driven outflows from luminous accreting supermassive black holes are an important part of active galactic nucleus (AGN) feedback. The effective Eddington limit, based on absorption of radiation by dust, not electron scattering, is revealed in the plane of AGN absorption column density $N_{\mathrm{H}}$ as a function of Eddington fraction $\lambda_{\mathrm{Edd}} = L_{\mathrm{bol}}/L_{\mathrm{Edd}}$, where a lack of objects is seen in the  region where the effective limit is exceeded. Here, we conduct radiation simulation using the {\sc cloudy} code to deduce the radiative force applied onto dusty gas at the nucleus and compare to the gravitational force to reveal the outflow region and its boundary with long-lived absorption clouds. We also investigate how the outflow condition is affected by various AGN and dust properties and distribution. As expected, the dust abundance has the largest effect on the \nhl diagram since the higher the abundance, the more effective the radiative feedback, while the impact of the inner radius of the dusty gas shell, the shell width and the AGN spectral shape are relatively negligible. The presence of other central masses, such as a nuclear star cluster, can also make the feedback less effective. The AGN spectral energy distribution  depends on the mass of the black hole and its spin. Though the effects of the AGN SED on the diagram are relatively small, the fraction of ionizing ultraviolet (UV) photons from the blackbody accretion disc is affected more by black hole mass than spin, and can influence the efficiency of radiation pressure.
\end{abstract}





\section{Introduction}
We have known for almost two decades that essentially all normal massive galaxies contain at their centers massive black holes, and furthermore, it appears that the mass of the black hole is strongly correlated with the stellar mass measured within the galaxy, the mass ratio being roughly $1:1000$ \citep{2001AIPC..586..363K,	2011MNRAS.413.1479S, 2002MNRAS.331..795M}. The major problem in astrophysics follows: how can the tiny black hole couple with large host galaxy when the ratio of the size of the black hole and that of its host galaxy is one to more than one hundred million?

It is believed that the co-evolution of these quite distinct components is due to feedback \citep{1998AA...331L...1S} \citep[for radiative feedback, see][]{1999MNRAS.308L..39F, 2005ApJ...618..569M, 2006MNRAS.373L..16F, 2008MNRAS.385L..43F, 2009MNRAS.394L..89F, 2012ARAA..50..455F,  2015MNRAS.449..147T, 2018MNRAS.481.4522I}, i.e. the fact that when mass is added to the central black hole it will emit light and high-velocity winds that penetrate into the surrounding galaxy, heat it, and expel matter from it, thus regulating the mass that can accumulate in the galaxy. Moreover, it is thought that feedback from black hole accretion eventually shuts off star formation in the galaxy, but details of the mechanism are still unclear.
In this paper we explore the  role of radiation pressure acting on dusty gas around black holes.

Many studies \citep[e.g.][]{2011ApJ...728...58B, 2014ApJ...786..104U, 2015ApJ...815L..13R} show that massive black holes are obscured by large columns of gas and dust within a few to tens of parsecs around black holes, and whether they are obscured or not depends on the accretion rate onto the central black hole.  While quasars (QSOs) appear to be basically at the Eddington limit $L_{\mathrm{Edd}}$ which is related to the Thomson scattering cross section $\sigma_{T}$ on electrons, expressed as 
\begin{equation}
	L_{\mathrm{Edd}}=\frac{4\pi G c m_{\mathrm{p}}M_{\mathrm{BH}}}{\sigma_{T}},
\end{equation}
where $c$, $m_{\mathrm{p}}$ and \mbh are speed of light, the proton mass and the black hole mass respectively, dust grains are partially charged in QSO, which is instead connected to the equivalent dust cross section per proton $\sigma_{\mathrm{d}}$ which is about 1000 times higher than the Thomson value: $\sigma_{\mathrm{d}}/\sigma_{T}\sim 1000$ \citep{2008MNRAS.385L..43F} where photoelectric opacity and line driving are important.

Thus, it means that a QSO at the standard Eddington limit for ionized gas is at the effective Eddington limit $L_{\mathrm{Edd}}^{\mathrm{eff}}$ for dusty gas in the galaxy considering the incredible correlation that mass of galaxy to mass of black hole is about the same: $M_{\mathrm{gal}}/M_{\mathrm{BH}}\sim 1000$. Here, the effective Eddington limit is defined as the limit when radiation pressure ejects mass based on absorption of radiation by dust, not on electron scattering. It also shows that radiation pressure by an active galactic nucleus (AGN) acting on such dusty gas can push and drive material outwards, appearing as outflows.

The effective Eddington limit is revealed in the plane of absorption column density $N_{\mathrm{H}}$ as a function of Eddington fraction $\lambda_{\mathrm{Edd}} = L_{\mathrm{bol}}/L_{\mathrm{Edd}}$, where a lack of objects is seen in the super-Eddington outflow region, in the range $10^{22} - 10^{25}$ cm$^{-2}$ and $\lambda>0.05$, compared with many AGNs in the long-lived absorption region \citep{2008MNRAS.385L..43F, 2010MNRAS.408.1714R}. The diagram became more complete with the work of \citet{2017Natur.549..488R} using the Swift Burst Alert Telescope (BAT) survey as 392 AGNs are contained in the diagram.

The \nhl diagram displays the so-called forbidden region where any AGNs in the region have outflows provided that the gas is close enough to the AGN since the black hole mass can dominate over more mass components mainly located away from the galactic center such as dark matter and stars which potentially hold the gas within the galaxy, so any radiation-absorbing materials near an AGN could be pushed out as an outflow.

Therefore, radiation pressure has an important role in determining how much gas can be close to  an AGN. It is crucial to reveal more precisely the outflow boundary curve in the \nhl plane with different dust properties: grain  abundance including dust-gas ratio, dust distribution and so on as well as the AGN properties: mass of the central black hole, its bolometric luminosity and AGN spectrum.

On a theoretical side, \citet{2008MNRAS.385L..43F} initially studied the radiation pressure on dusty absorbing gas followed by a simple 1D simulation of radiation pushing molecular clouds being conducted in \citep{2012MNRAS.427.2998I}. Then, \citet{2015MNRAS.449..147T} investigated whether the gas clouds which are optically thick, particularly in infrared (IR), are relevant. If most gas initially resides within inner 100pc then it can be optically thick to reprocessed IR, so the radiation pressure and hence the momentum can be boosted by $(1+\tau_{IR})$ due to IR radiation trapping, where $\tau_{IR}$ is the optical depth in IR. In a dusty spherical geometry the light will shift to longer wavelength as it works its way out.  It is then less effective at momentum transfer due to the smaller cross section. If $\tau_{IR}$ is few, the momentum of the outflowing gas could be expressed as $\dot M v\sim10 (L/c)$ \citep{2015MNRAS.451...93I, 2018MNRAS.479.2079C}. However, it is still unclear how the dust and AGN properties influence outflow occurrence.

Here, using the radiation simulations, we model typical AGN radiation-pressure driven outflows with initial obscuration, to see their trajectories on the $N_{\mathrm{H}} - \lambda_{\mathrm{Edd}}$ plane and investigate how the outflow condition is affected due to various AGN and dust properties and distribution. 

This paper is structured as follows: the radiation simulations being conducted and their purpose are described in Section~\ref{sec:sim}. The results of simulations as the effects of AGN and dust properties in the \nhl diagram are shown in Section~\ref{sec:results}. Finally, the results are summarized and discussed in Section~\ref{sec:discussion}.

\section{Radiation simulation}
\label{sec:sim}
\subsection{Radiation code and scientific purpose}
\label{sec:cloudy}
We use the {\sc cloudy} photoionization simulation code \citep{2017RMxAA..53..385F}, which determines radiative force so that it can be tested as a function of various parameters such as the distance between the cloud and the center of the central object (i.e. radius), that between the illuminated face of the cloud and a point within the cloud (i.e. depth) and column density. We utilize {\sc cloudy} simulations to estimate the following.

\begin{enumerate}
	\item Effects of dust properties and abundance on radiation pressure.\\
		Principally dust grains absorb and scatter radiation which has a comparable wavelength to their size.
	\item Effects of black hole mass and spin on the AGN spectrum.\\
			Most of the emission from accretion comes out in the ultraviolet (UV) from the accretion disk. The more massive a black hole is, the larger both the black hole and emitting region are, as the Schwarzschild radius is given as \rs $=2GM_{\mathrm{BH}}/c^2 \propto M_{\mathrm{BH}}$, and the size of accretion disks can be expressed as $r \propto M_{\mathrm{BH}}^{2/3}$ from standard thin-disk theory \citep{1973AA....24..337S}. For black-body radiation, from standard accretion theory, the peak temperature of the
accretion disc $T_{\mathrm{max}}$ decreases with increasing black hole mass as $T_{\mathrm{max}} = (3G \dot M M_{	\mathrm{BH}} /8\pi \sigma R^3 )^{1/4}$ where $\sigma$ is Stefan–Boltzmann constant. As the fraction of the energy coming out at low energy, shifted to the optical band, increases, the coupling between radiation and dust is reduced. For the most massive black holes, this means that radiation pressure and hence outflows may be less effective.
	\item Effects of the AGN spectrum on the radiative force on the dusty gas.\\
			Radiation pressure on dust in particular depends on the optical-UV-EUV spectrum of the AGN.
\end{enumerate}

For the radiation simulations, we assume a single constant-density gas and dust shell surrounding a central black hole in an AGN is distributed spherically symmetric, and hence closed geometry is assumed.
 Gas and dust particles could be weakly ionised e.g. due to X-ray emission and cosmic rays from
the AGN, and then dust and gas electrically couple by Coulomb forces.

We also assume the dust shell works ideally as a rigid body as the materials are magnetically coupled so that we can simply compare radiative force with gravitational force applied onto the dust shell in order to reveal the condition of occurrence of outflows. The radiation fields which enter in the calculations are mainly the following: incident radiation, reflected radiation at the illuminated face of the dust, diffusion inside the shell and transmission from the shielded outer face of the dust.

In {\sc cloudy} simulations, the following parameters are relevant: number density of protons in the dust shell $n_{\mathrm{H}}$, the inner radius of the dust shell \rin as the dust sublimation radius, the shell width \rsh, the abundance of the dust $Z$ and the bolometric luminosity \lbol of an AGN and AGN spectrum. The column density \nh of the gas can be calculated as \nh $=n_{\mathrm{H}}$\rsh. In this study, we use \rin$=5$ pc, \rsh$=1$ pc and \lbol$=10^{45}$ erg/s as a standard parameter set, unless otherwise noted. Raising the hydrogen density means increasing the column density within a given radius and also increasing the recycling of the radiation field inside the shell.

The {\sc cloudy} simulation with the grain code ISM and Orion applies the default abundance where the gas-phase abundance of the shell is equivalent to that of the ISM of our galaxy, with the graphite and silicate grain abundance and size distribution  set to the values along the line of sight to the Trapezium stars in Orion. The dust to gas ratio and grain to metal ratio are also fitted to the Orion values. $Z=1.0$ indicates the gas-phase and grain abundance described above. For instance, $Z=0.5$ means both of these abundance values are halved. The Orion grains are large, fairly gray, grains. The ISM grains have a broader size distribution and will absorb UV more efficiently. The dust opacity of the molecular clouds can be estimated from the grain abundance. A comparison between the Orion model and ISM model is shown in Section \ref{sec:grain}.

\subsection{Mass-weighted mean acceleration vs. mean radiative acceleration due to radiation pressure on dust}
\label{sec:balance}

The gravitational acceleration on the dust shell surrounding the central black hole can be estimated as
\begin{equation}
	a_{\mathrm{grav}}(r)\sim\frac{G M_{\mathrm{BH}}}{r^2},
\end{equation}
where $G$ is the gravitational constant, and \mbh is mass of the central black hole of the AGN.

The gravitational force on the dusty gas is expressed in terms of the proton mass $m_{\mathrm{p}}$ as
\begin{equation}
	F_{\mathrm{grav}}(r)=4 \pi r^2 n_{\mathrm{H}} m_{\mathrm{p}} a_{\mathrm{grav}}(r) dr,
\end{equation}
and the gravitational net force is given by
\begin{align}
	F_{\mathrm{grav, net}}&=\int F_{\mathrm{grav}}(r)\\
	&=4 \pi G M_{\mathrm{BH}} n_{\mathrm{H}} m_{\mathrm{p}} R_{\mathrm{sh}}.
\end{align}

Then, the mass-weighted mean gravitational acceleration on a dust shell is given by
\begin{equation}
	\langle a_{\mathrm{grav}} \rangle = \frac{F_{\mathrm{grav, net}}}{M_{\mathrm{tot}}},
\end{equation}

where $M_{\mathrm{tot}}$ is the total mass of the dust shell expressed with the outer radius $R_{\mathrm{out}} = R_{\mathrm{in}} +R_{\mathrm{sh}} $ as:

\begin{equation}
	M_{\mathrm{tot}}=\int _{R_{\mathrm{in}}}^{R_{\mathrm{out}}} 4\pi r^2 n_{\mathrm{H}} m_{\mathrm{p}} dr.
\end{equation}

Since the simulations show the radiative acceleration \ar, the mean radiative acceleration \arm of the dust shell can be estimated. Then, a balance between \arm and \agm is revealed so that the balancing curve \arm$=$\,\agm, which represents the effective Eddington limit, can be plotted on the \nhl diagram to estimate the outflow region. In this paper, as a possible additional central gravitational component besides the black hole, we  consider the presence of a nuclear star cluster (NSC) \citep[see review by][]{Neumayer2020}. We investigate its effect where it increases gravitational force on the dusty gas by the presence of NSC in the vicinity of the central black hole within the inner radius of the dust shell, as discussed in Section \ref{sec:abun}.

\section{Results of the simulations}
\label{sec:results}
\subsection{Effects of dust abundance and other gravitational factors on \nhl diagram}
\label{sec:abun}

\begin{figure}
	\includegraphics[width=\columnwidth]{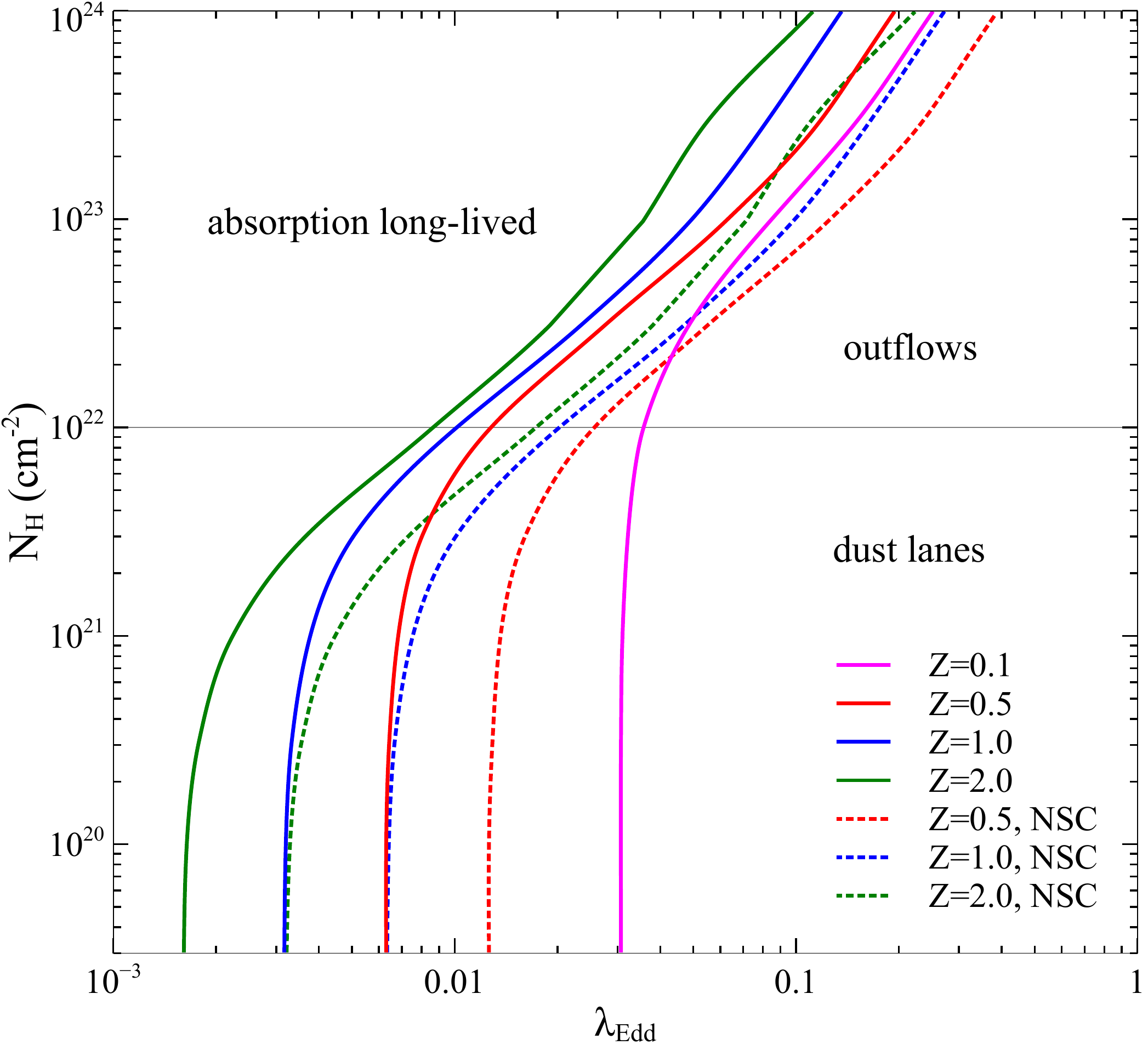}
    \caption{The \nhl diagram with various abundance values $Z$ with and without NSC effects being taken into account: $Z=0.1$ (magenta), $Z=0.5$ (red), $Z=1.0$ (blue), $Z=2.0$ (green) excluding nuclear star cluster (NSC) (solid) and including NSC (dot). The horizontal line corresponds to \nh$=10^{22}$\cms (grey solid). For dotted curves, it is assumed that the NSC, the mass of which is similar to \mbh, is distributed in the vicinity of the central black hole.}
    \label{fig:abun}
\end{figure}

The effects of abundance of the dust shell on the \nhl plane are shown in Fig.~\ref{fig:abun} where the various abundance values $Z$ with and without NSC effects, the increase of gravitational force on the dusty gas by the presence of NSC as introduced in Section \ref{sec:balance}, are taken into account: $Z=0.1$ (magenta), $Z=0.5$ (red), $Z=1.0$ (blue), $Z=2.0$ (green) excluding NSC (solid) and including NSC (dot). Here we display $Z=0.1$ model to approximately estimate the limit for the case of "no grains" model so that this can be compared with the other grain models. As noted in Section~\ref{sec:cloudy}, other parameters are set to \rin$=5$ pc, \rsh$=1$ pc and \lbol$=10^{45}$ erg/s. For these comparisons, an AGN SED which is available with the command "table AGN" in the {\sc cloudy} code is commonly used. Its continuum is similar to typical radio quiet active galaxies deduced by \citet{1987ApJ...323..456M}, the detail of which is described in \citep{2006hbic.book.....F}. 

The horizontal grey solid line corresponding to \nh$=10^{22}$\cms\ shows where outer dust lanes from the host galaxy would contribute to the line of sight column density, which means the absorption by the outer dust lanes would be important for \nh$<10^{22}$\cms. (Distant shells of gas would require prohibitive mass for the given high column density \citep{2008MNRAS.385L..43F, 	2009MNRAS.394L..89F}). So, this line represents dusty gas further out in the galaxy, such as the dust lanes seen in MCG-6-30-15. We stress that the results are all line-of-sight and mean that if the central emission is anisotropic, corrections are necessary to go from that luminosity to the true bolometric luminosity of the galaxy. So the Eddington fraction in our diagrams is not the same as an angle averaged one as discussed in \citet{2019MNRAS.486.2210I}.

The Compton thickness can be expressed as \nh$\sim1/$\com$\sim10^{24}$\cms in the optical and UV if the gas remains ionized, as well as for all physical states for hard gamma rays. The Compton-thick gas clouds, \nh$>10^{24}$\cms, can be long-lived provided that \lam$<1$ and might be likely to experience star formation \citep{2008MNRAS.385L..43F, 	2009MNRAS.394L..89F}. Hence, the diagram focuses on displaying only Compton-thin AGNs within \nh$<10^{24}$\cms.

The left side of the effective Eddington limit curve of \arm$=$\,\agm shows the region where long-lived absorbing clouds can survive as gravity overcomes radiation pressure, while the right side represents the outflow region, also called the forbidden or blowout region. For \nh$>10^{22}$\cms as the radiative force dominates and drive away the obscuring materials to see the nucleus with the absorption being only variable or transient. Most of the UV photons from the AGN are used up within the innermost $10^{22}$\cms of a shell. The outer dust shell acts as dead weight and is pushed by the inner shell which was accelerated by the radiation pressure. The outflow region also suggests that even at \lam$\ll1$, radiative feedback can clear out the nearby obscuring materials around the nucleus. Thus, though it depends on viewing angle whether the object is obscured or unobscured, generally obscured AGN reside to the left and in the dust lanes region while unobscured AGN exist in the outflow region and below \nh$=10^{20}$\cms in the diagram \citep{2017Natur.549..488R}. Fig.~\ref{fig:abun} is also basically analogous to Figure 3 in \citet{2018MNRAS.479.3335I} where it is discussed why the three curves tend to become less distinguishable in the single scattering regime (around $\log N_H \sim22-23$), and the {\sc cloudy} simulations  corroborate the analytic results. The effect of the enclosed mass was also previously noted by e.g. \citet{2009MNRAS.394L..89F}.

If the abundance is raised, it induces the increase of the opacity and thus the radiative acceleration. So less bolometric luminosity is needed to offset the gravitational force, which results in shifting the curve to the left on the diagram as shown in Fig.~\ref{fig:abun}.

The effects of any NSC are also demonstrated in Fig.~\ref{fig:abun} where we assume a mass for the NSC  is similar to \mbh, distributed in the vicinity of the central black hole within the inner radius of the dust shell. If there is such an NSC, it approximately doubles \agm. It means the required \arm and \lbol are doubled for the balance between \arm and \agm to be maintained, and therefore \lam is doubled, which results in the curve being shifted to the right on the diagram as shown in Fig.~\ref{fig:abun}. NSC are reviewed by \cite{Neumayer2020}, where it is shown that they are common in galaxies with stellar masses less than $10^{10}$\msun.

 It is shown that not only the abundance but other factors, such as NSC, than the central black hole which generates significant gravitational force on the dust have crucial impacts on the curve in the \nhl plane.

\subsection{Effects of dust shell width, inner radius and AGN bolometric luminosity on \nhl diagram}
\label{sec:other}
\begin{figure}
	\includegraphics[width=\columnwidth]{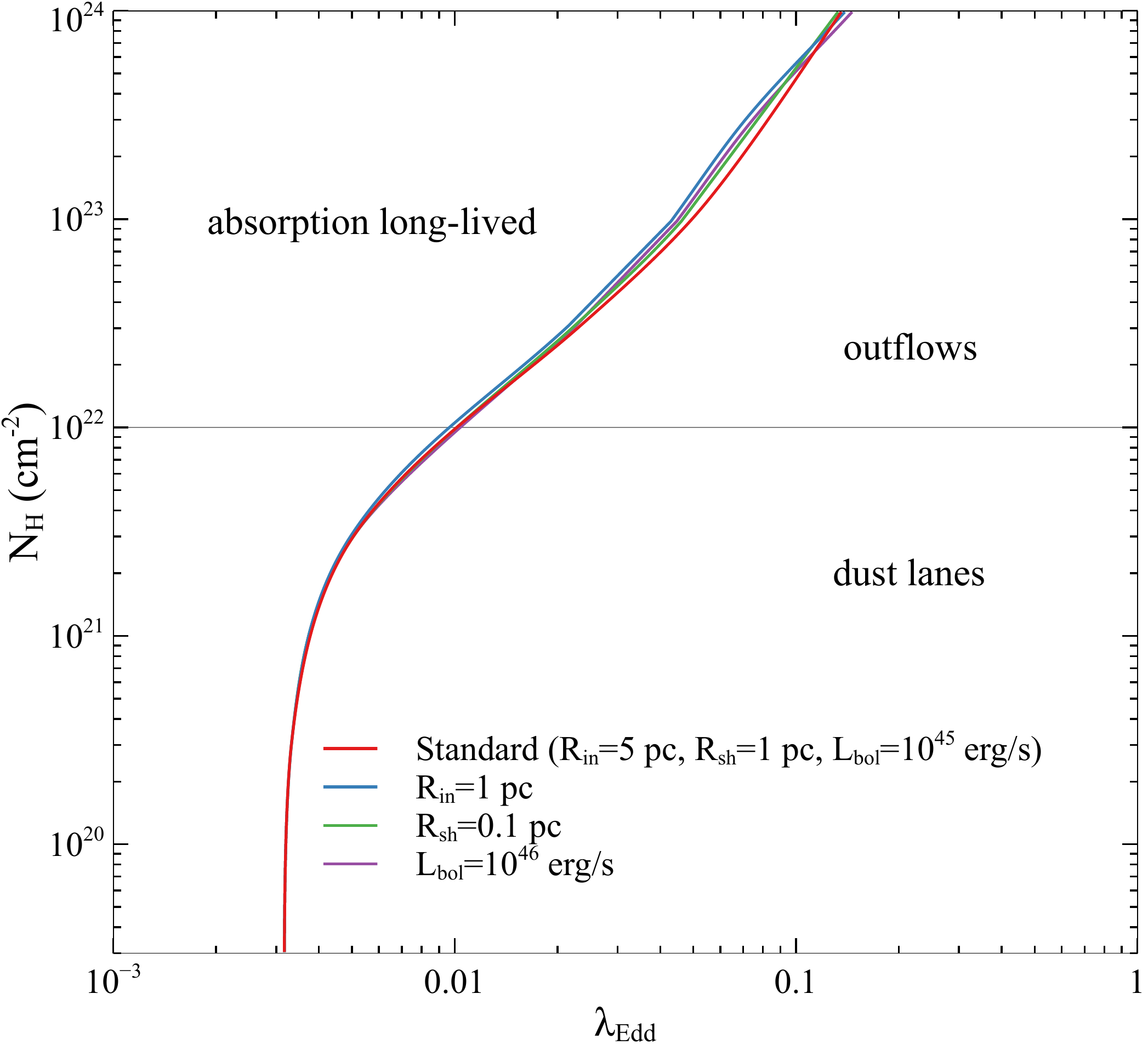}
    \caption{The dependence of the dust shell width, inner shell radius and bolometric luminosity on the curves on the \nhl diagram with abundance being fixed to $Z=1.0$ for this comparison: the standard model with \rin$=5$ pc, \rsh$=1$ pc and \lbol$=10^{45}$ erg/s (red), one with different \rin of $1$ pc (cyan), one with \rsh $=0.1$ pc (green) and one with \lbol$=10^{46}$ erg/s (purple). It is shown that the dependence on these factors are negligible.}
    \label{fig:other}
\end{figure}

Examples of the effects of the dust shell width, the inner radius of the shell the dust sublimation radius and the AGN bolometric luminosity are shown in Fig.~\ref{fig:other}. For these comparisons, abundance is fixed to $Z=1.0$. We show the standard model with \rin$=5$ pc, \rsh$=1$ pc and \lbol$=10^{45}$ erg/s and other three models where one of these parameters are changed: the model with \rin $=1$ pc, one with \rsh $=0.1$ pc and one with \lbol $=10^{46}$ erg/s.

As displayed, the dependence on those parameters are negligibly small, $< 5$\%, compared with that of abundance and NSC as shown in Section~\ref{sec:abun}.

\subsection{Effects of black hole mass and angular momentum on \nhl diagram}
\subsubsection{Effects of black hole mass and angular momentum on AGN spectrum}
\label{sec:sed}
As described in Section~\ref{sec:cloudy}, while black hole mass should affect the AGN SED (spectral energy distribution) we also investigate the effects of black hole angular momentum on the SED. Here we only consider disk black-body emission without taking into account coronal emission, spectral hardening or other factors as we assume it would be reasonable for explaining the spectrum given by a disk around a SMBH. SEDs with various sets of \mbh and dimensionless spin parameter $a_*$ are generated using {\sc kerrbb} \citep{2005ApJS..157..335L}, a model of a general relativistic accretion disc around a Kerr black hole with {\sc xspec} version $12.11.0$: \mbh $= 10^6$, $10^7$, $10^8$, $10^9$, $10^{10}$\msun and $a_*=-1$, $0$, $0.9$, $0.99$. The spin parameter is defined as $a_* = cJ/GM_{\mathrm{BH}}^2$ where $c$ is speed of light, and $J$ is the angular momentum of the black hole.

We also set some of the {\sc kerrbb} parameters as follows: the distance from the observer to the black hole is 5 pc as the value of \rin, and the effective mass accretion rate of the disk is calculated with the fixed \lam $=0.1$ for each model. The other parameters are set to the default values (e.g. disk inclination angle which is the angle between the axis of the disk and the line of sight is set to $i =30$; ratio of the disk power produced by a torque at the disk inner boundary to the disk power arising from accretion $\eta=0$ which represents a standard Keplerian disk with zero torque at the inner boundary.) Note that since we consider $\eta = 0$,  the effective mass accretion rate just corresponds to the mass accretion rate $\dot M$ of the disk.

\begin{figure*}
	\begin{minipage}{0.49\hsize}
	\includegraphics[width=\columnwidth]{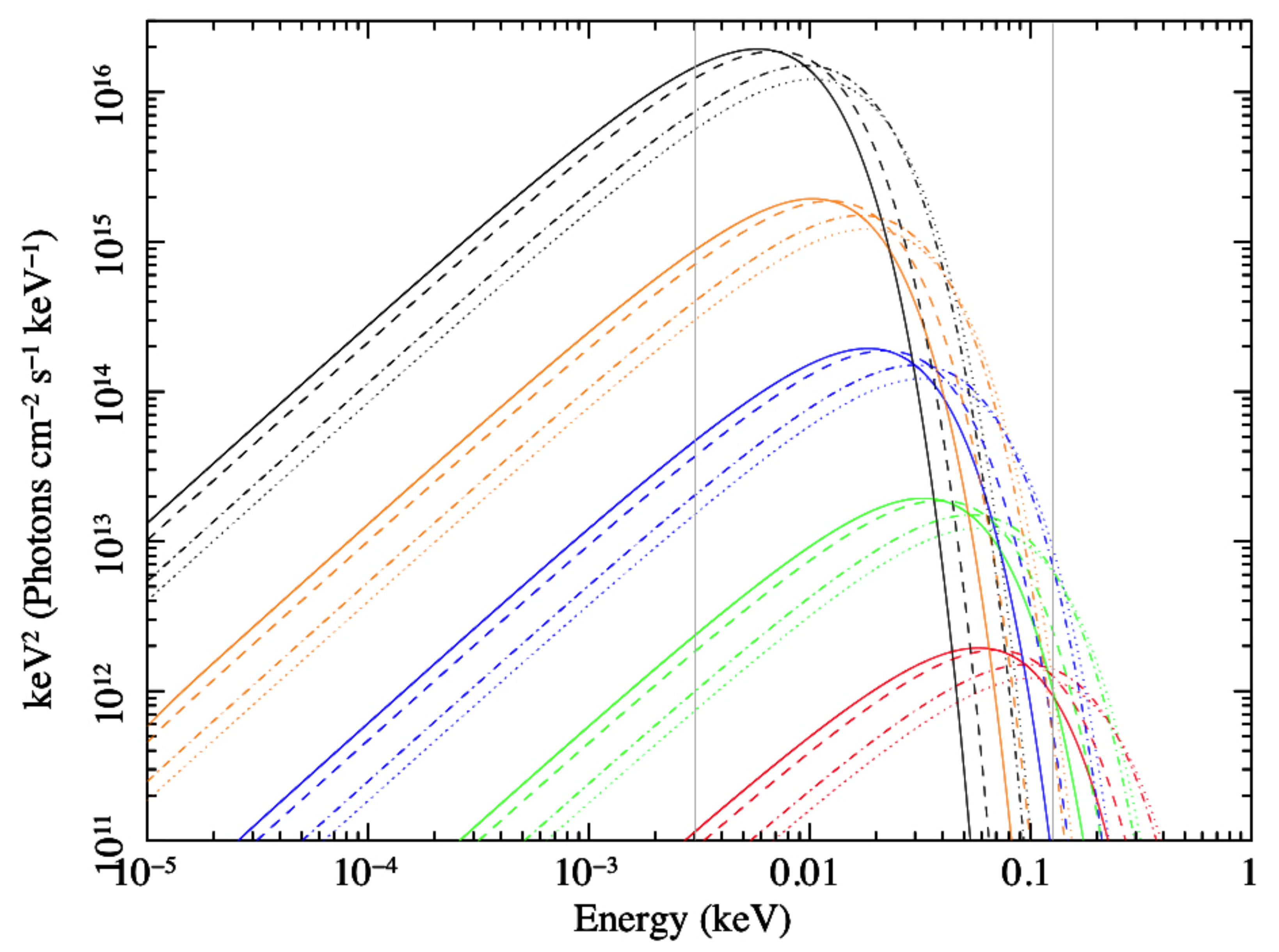}
\end{minipage}
\begin{minipage}{0.49\hsize}
	\includegraphics[width=\columnwidth]{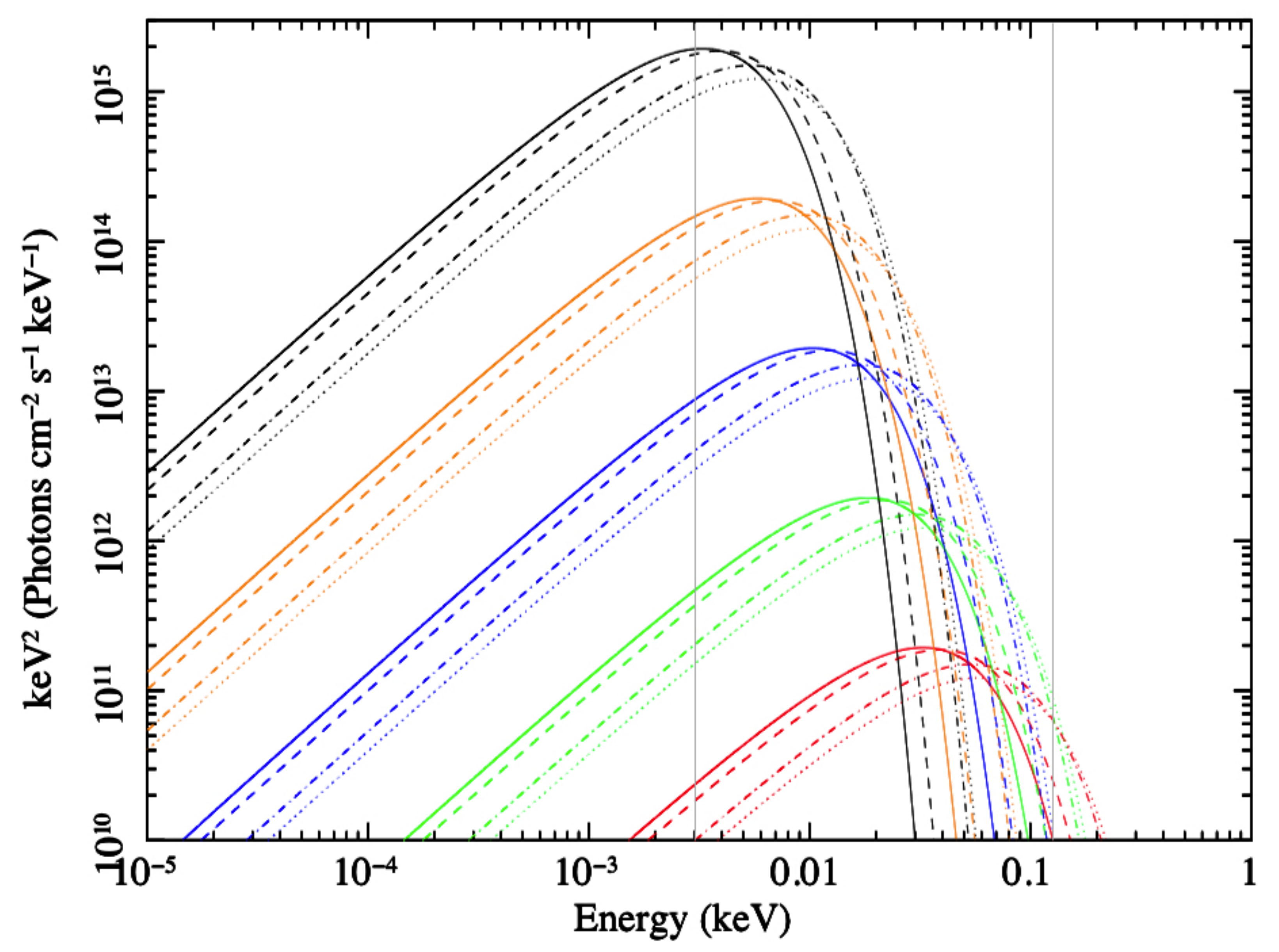}
\end{minipage}	
\begin{minipage}{0.49\hsize}
	\includegraphics[width=\columnwidth]{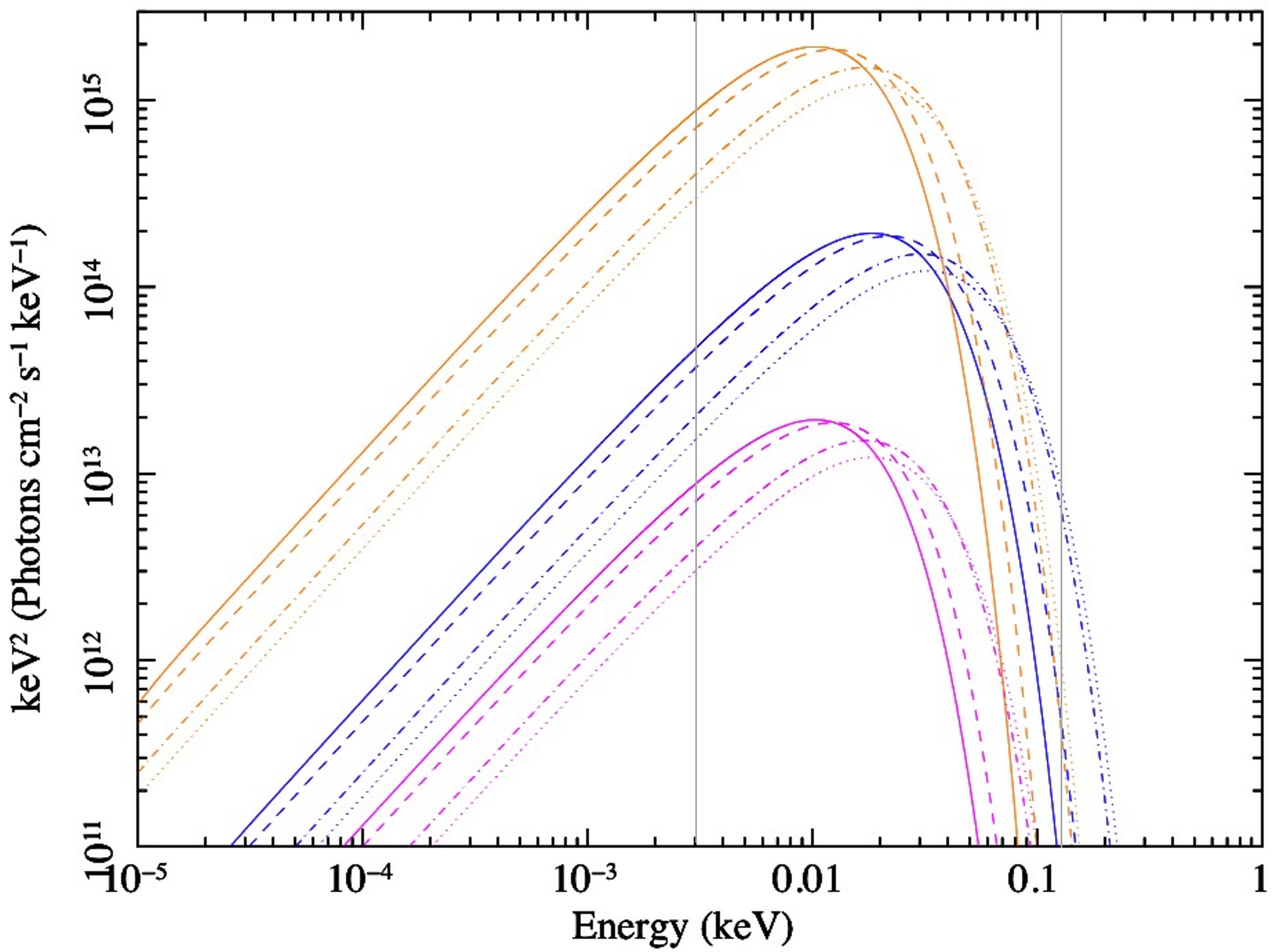}
\end{minipage}	
\vspace*{+0.5cm}
    \caption{(Top) AGN SEDs of fixed \lam $=0.1$ (left) and \lam $=0.01$ (right) with various parameter sets of central black hole mass \mbh and spin parameter $a_*$: \mbh $=10^6$\msun (red), \mbh $=10^7$\msun (green), \mbh $=10^8$\msun (blue), \mbh $=10^9$\msun (orange), \mbh $=10^{10}$\msun (black), $a_* = -1$ (solid), $a_* = 0$ (dash), $a_* = 0.9$ (dash-dot) and $a_* = 0.99$ (dot). Vertical grey lines represent the energy range of UV photons, $10-400$ nm.  Note that the vertical axis on the left is $10$ times higher than the right. (Bottom) Comparison between models with fixed \lam $=0.1$ and those with \lam $=0.01$: for \lam $=0.1$,  \mbh $=10^8$\msun (blue) and \mbh $=10^9$\msun (orange); for \lam $=0.01$, \mbh $=10^8$\msun (magenta). Likewise, $a_* = -1$ (solid), $a_* = 0$ (dash), $a_* = 0.9$ (dash-dot) and $a_* = 0.99$ (dot).}
    \label{fig:sed}
\end{figure*}

While the bolometric luminosity is expressed as \lbol $=$ \lam \led $=\epsilon \dot M c^2$ where $\epsilon$ is the accretion efficiency, note that the efficiency depends on spin parameter. Here, for the creation of the Kerr black hole spectra, we take the accretion efficiency $\epsilon=0.038$ for $a_*=-1$, $\epsilon=0.057$ for $a_*=0$, $\epsilon=0.16$ for $a_*=0.9$ and $\epsilon=0.26$ for $a_*=0.99$ from \citet{1973grav.book.....M}. The inner most circular orbit (ISCO) radius of the disk is \risco $=4.5$ \rs, $3$ \rs, $1.16$ \rs and $0.72$ \rs for $a_*=-1, 0, 0.9$ and $0.99$ respectively, where \rs is the Schwarzschild radius \rs $=2GM_{\mathrm{BH}}/c^2$. The resultant SEDs with \lam being fixed to $0.1$ and $0.01$ are shown on the top in Fig.~\ref{fig:sed}. On the bottom, the comparison between them is displayed.

The results show if the spin is higher, the energy becomes higher as the spectrum is shifted to the higher energy while the black hole mass shows the opposite trend: more massive black hole causes the shift to lower energy. As explained in Section~\ref{sec:cloudy}, it is due to black-body radiation. If \lam and spin are fixed, the Newtonian black-body radiation theory on scaling law of effective temperature \tmax $\propto$ \mbh $^{-1/4}$ is well fitted with the results (comparison between blue and orange). The scaling of \tmax $\propto$ \lbol $^{1/4}$ \risco $^{-1/2}$ can also be expressed as \tmax $\propto$ \lbol$^{1/4}$\mbh$^{-1/2}\propto$ \lam$^{1/4}$\mbh$^{-1/4}$ with a fixed spin due to \risco $\propto$ \rs $\propto$ \mbh. Therefore, \tmax of a model with \lam $=0.1$ and \mbh = $10^9$ \msun (orange) is similar to that of \lam $=0.01$ and \mbh = $10^8$ \msun (magenta) with a spin fixed. On the other hand, if \lam and \mbh are fixed, it means \lbol and \rs become constant, so \tmax $\propto$ \risco $^{-1/2}$. The degeneracy between mass, accretion rate and spin is also discussed in \citet{2018AA...612A..59C}.

Another important point is as \lam increases, the SED shifts to  higher energy. As the fraction of the energy coming out at high energy, shifted to the UV from the optical band, increases the coupling between radiation and dust could be increased.

\subsubsection{Effects of AGN SED on radiative force}
\label{sec:accel} 
If the fraction of UV photons applied onto the dust decreases, the coupling between radiation and dust could be reduced. We try to see if it could make the radiation pressure less effective.

\begin{figure*}
\begin{minipage}{0.49\hsize}
	\includegraphics[width=\columnwidth]{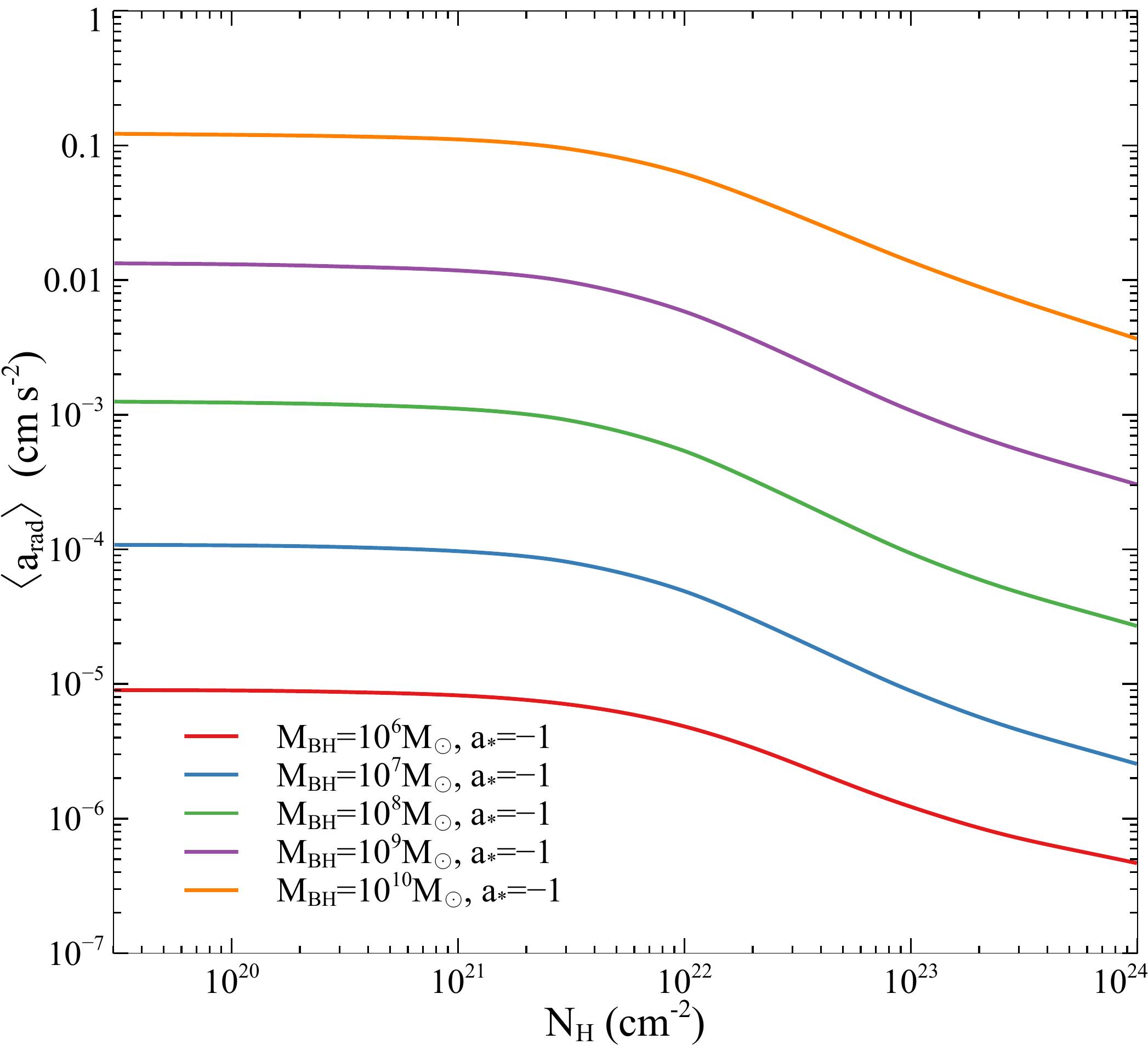}
\end{minipage}
\begin{minipage}{0.49\hsize}
	\includegraphics[width=\columnwidth]{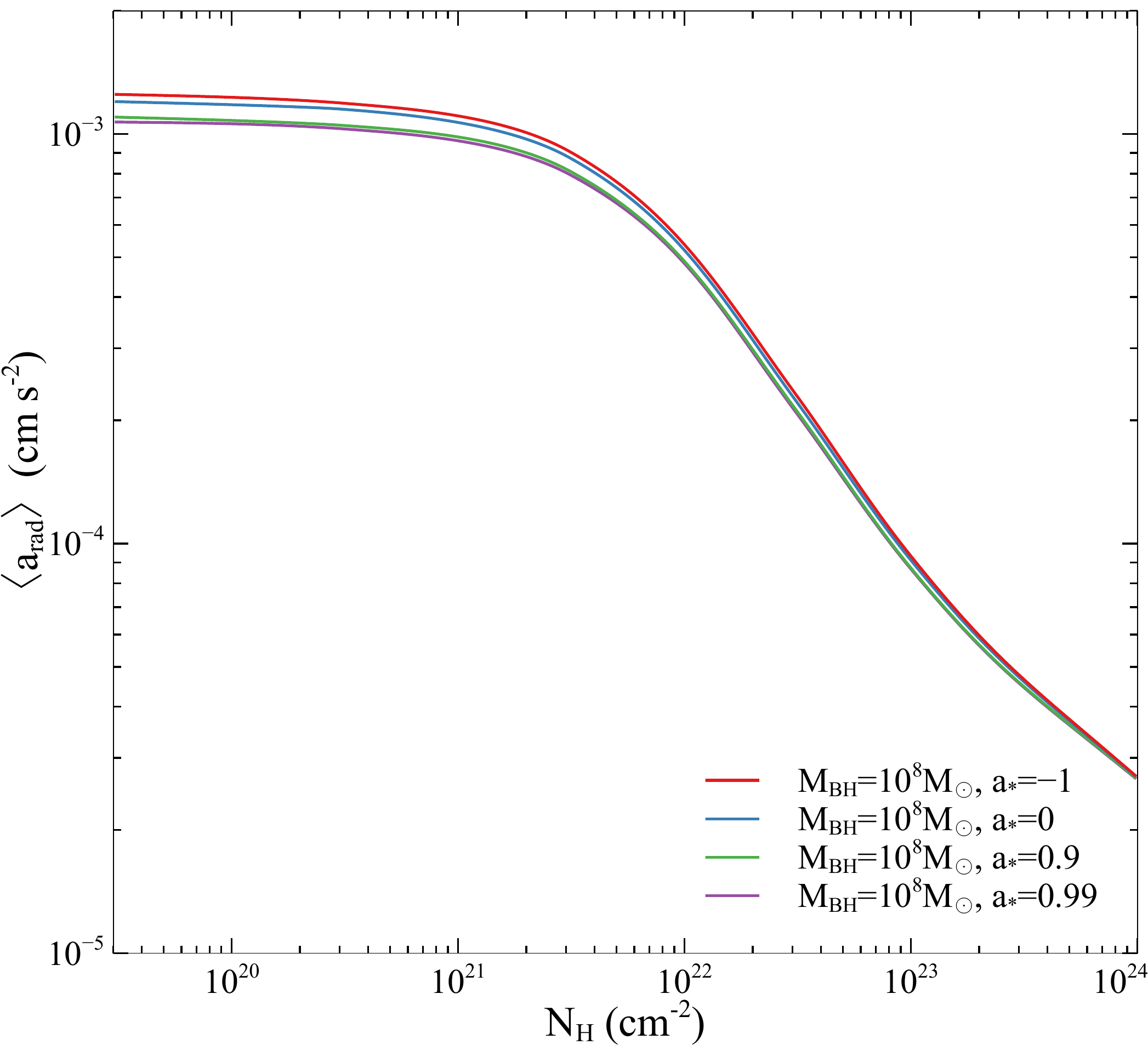}
\end{minipage}
\vspace*{+0.5cm}
\caption{(Left) The \anh diagram with a fixed spin parameter $a_* = -1$ and different black hole masses: \mbh $=10^6$\msun (red), \mbh $=10^7$\msun (cyan), \mbh $=10^8$\msun (green), \mbh $=10^9$\msun (purple) and \mbh $=10^{10}$\msun (orange). (Right) The \anh diagram with a fixed black hole mass \mbh $=10^8$\msun and different spin parameters: $a_* = -1$ (red), $a_* = 0$ (cyan), $a_* = 0.9$ (green) and $a_* = 0.99$ (purple).}
 \label{fig:anh}
\end{figure*}

The effects of those AGN SEDs on the radiative acceleration of the dust shell due to the radiation pressure are shown in Fig.~\ref{fig:anh}. Except AGN SEDs, all the simulations are conducted with a fixed abundance $Z=0.5$ and a standard parameter set: \rin$=5$ pc, \rsh$=1$ pc and \lbol$=10^{45}$ erg/s. The left figure is the \anh diagram with a fixed spin parameter $a_* = -1$ and the five black hole masses: \mbh $=10^6$\msun, \mbh $=10^7$\msun, \mbh $=10^8$\msun, \mbh $=10^9$\msun and \mbh $=10^{10}$\msun. The relation \arm$\propto$ \mbh is clearly shown if \lam is set to be constant so that \lbol can be expressed as \lbol$\propto$ \led$\propto$ \mbh.

The drop in acceleration when \nh exceeds $10^{22}$\cms is explained as due to the depletion of UV photons in the outer part of the dusty shell. As described in Section~\ref{sec:abun}, most of the UV photons are depleted by dust absorption within the inner $10^{22}$\cms. The outer dust shell thus serves only as a dead weight that decelerates the outflowing shell.

The physical trends with column density could also be interpreted in terms of simple analytic scalings. In the UV regime ($N\lesssim10^{22} \mathrm{cm}^{-2}$), \ar$\sim(\kappa_{\mathrm{UV}} L)/(4 \pi c r^2)$ where $\kappa_{\mathrm{UV}}$ is a UV dust opacity, i.e. the radiative acceleration is independent of the shell column density; while in the single scattering limit, the radiative acceleration (\ar$\propto L/(c
M_{\mathrm{sh}})$) decreases with increasing column.

On the other hand, the right-hand panel shows the effects of those four spin parameters with a fixed black hole mass \mbh $=10^8$\msun: $a_* = -1$, $a_* = 0$, $a_* = 0.9$ and $a_* = 0.99$. It is shown the higher the spin is, the less effective the radiation pressure is though there is merely a slight difference as the SED difference depending on spin variation shown in Fig.~\ref{fig:sed} is also small. This could be because the SED for higher spin contains slightly less fraction of UV photons even though the effective temperature increases as the spin becomes higher.

\subsubsection{Effects of AGN SED on \nhl diagram}
\label{sec:sed-nhl}

The effects of the five different black hole masses on \nhl diagram are shown in Fig.~\ref{fig:mbh}. As well as Section~\ref{sec:accel}, except AGN SEDs, all the simulations are conducted with $Z=0.5$ and a standrad parameter set: \rin$=5$ pc, \rsh$=1$ pc and \lbol$=10^{45}$ erg/s. Note that for generating the curves, SEDs based on the corresponding \mbh and spin values in the {\sc cloudy} code are rescaled depending on $\lambda_{\rm Edd}$ with the scaling of the energy of the maximum on flux profiles, $E_{\rm max} \propto L^{1/4}$, for a given spin as $\epsilon$ only depends on $r_{\rm ISCO}$ which in turn depends only on spin, and $T_{\rm max}$ should scale with $E_{\rm max}$ as shown in Section \ref{sec:sed}.

It is shown that the higher the spin is, the larger difference the black hole mass can induce. As \lam increases, SED shifts to the higher energy. As the fraction of the energy coming out at high energy, shifted to the UV from the optical band, increases, the coupling between radiation and dust could be increased. The reason why the slope in the \nhl plane for \lam $=0.01-0.1$ is getting steeper as \lam increases can be because the increase of the UV fraction makes the radiation pressure more effective which means the outflow would require lower \lam.

However, the common trend among the five \mbh models that the more massive the black hole is, the lower the UV fraction is in their SEDs, shows the inconsistency with the results in the \nhl diagram. The diagram instead shows for \nh $>10^{22}$ \cms, except the case of \mbh$=10^6$ \msun, the more massive the black hole is, the more effective the radiation pressure is.

The UV fraction changes as \lam changes since the SED changes being shifted to the lower energy as \lam decreases. So the UV fraction would be expressed as a function of \lam for each model. However, after all, it seems all models, if including soft X-ray fraction for \mbh $=10^6$ \msun model though, at least show the more massive $-$ the more UV (\& X-ray) fraction trend for \lam $=0.01-0.1$. Since the first optical depth reprocesses most of the EUV and XUV into optical and IR, which is then further processed into the IR, there would be differences in that first optical depth but not at depth.

The effects of the four spin parameters on the diagram are also shown in Fig.~\ref{fig:spin}. It is shown the more massive the black hole is, the smaller the difference spin can bring. However, overall, the spin dependence on the diagram is minor.

\subsection{Effects of grain properties on the spectrum and \nhl diagram}
\label{sec:grain}

\begin{figure*}
\begin{minipage}{0.49\hsize}
	\includegraphics[width=\columnwidth]{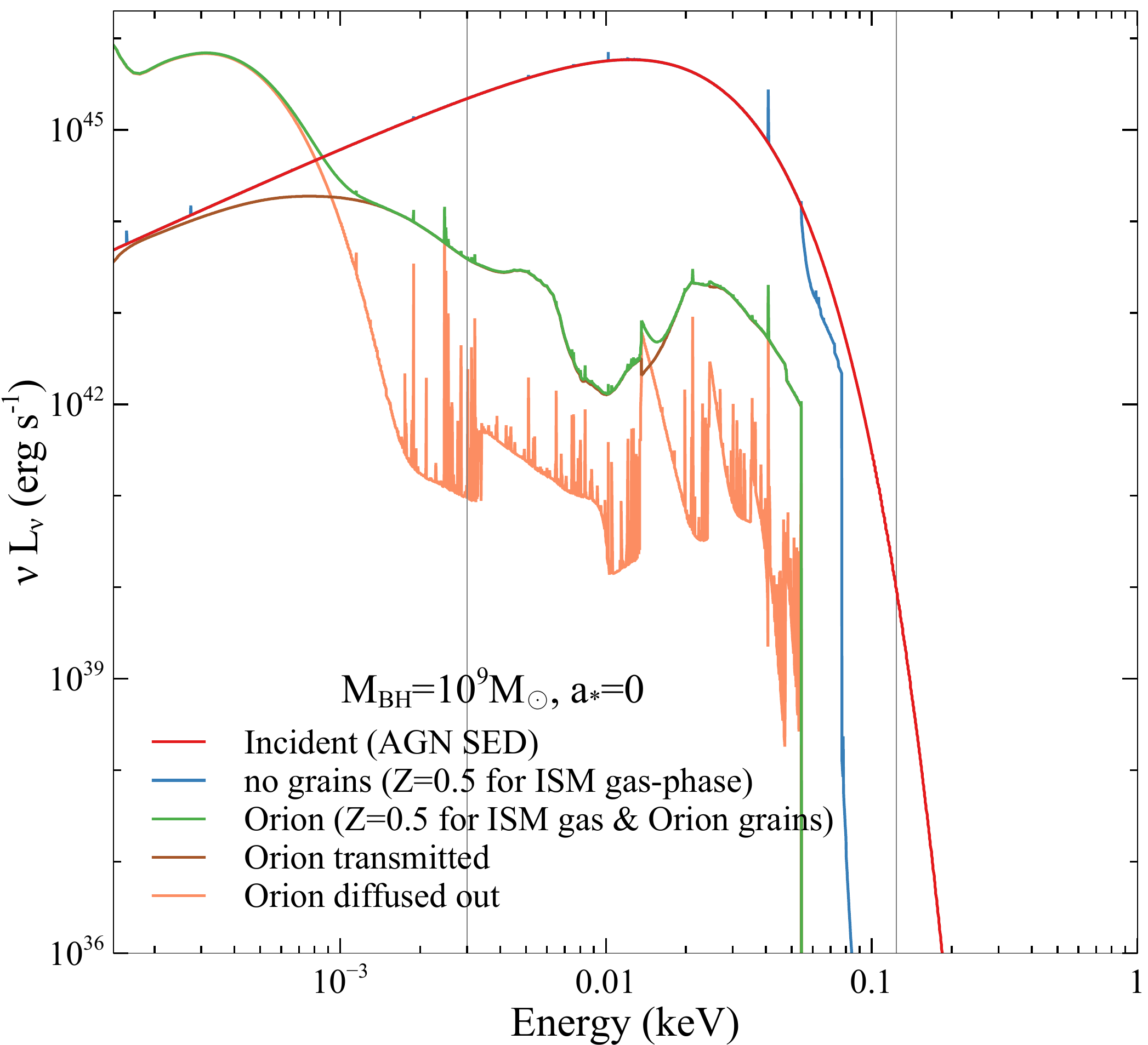}
\end{minipage}
\begin{minipage}{0.49\hsize}
	\includegraphics[width=\columnwidth]{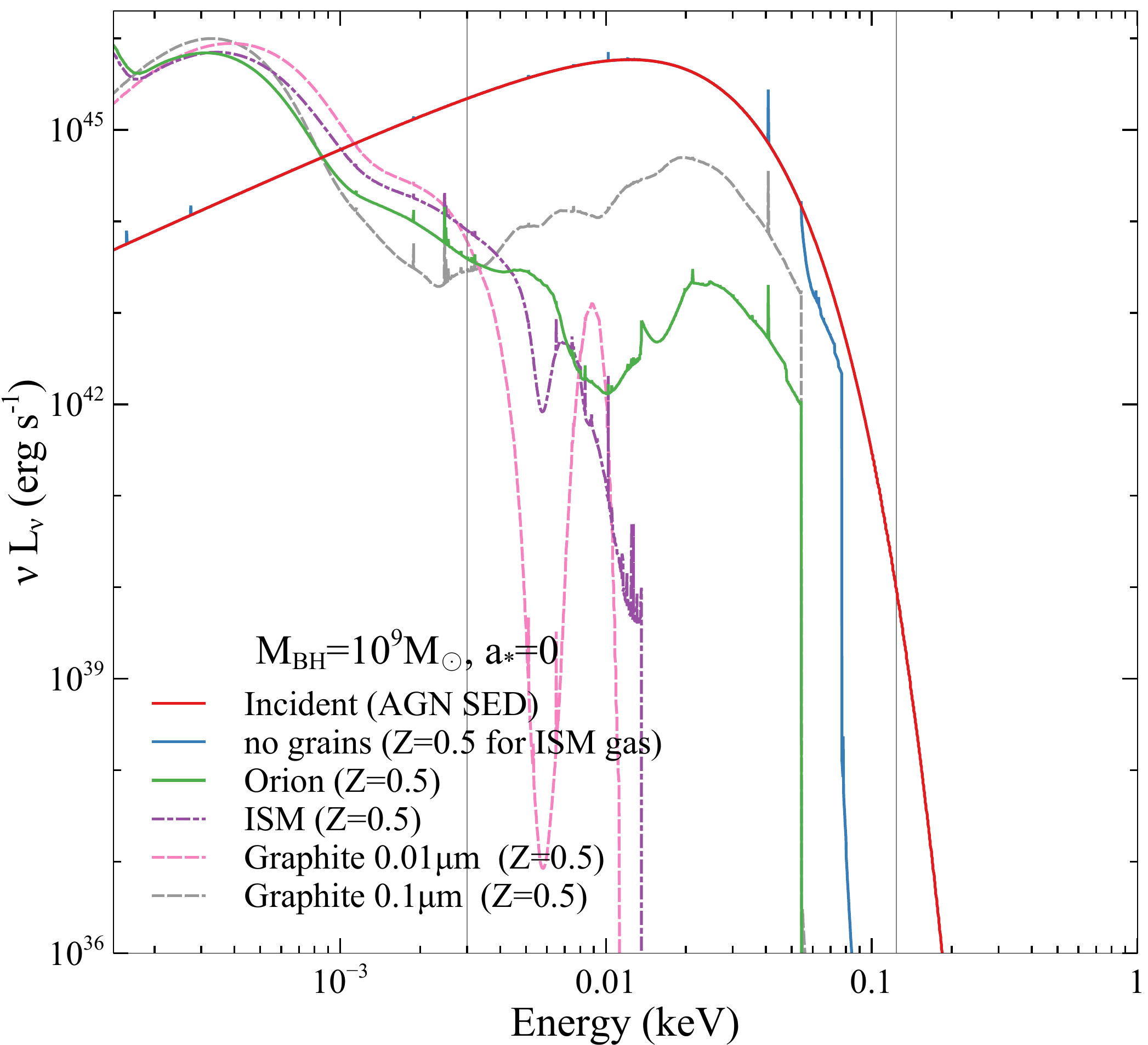}
\end{minipage}
\begin{minipage}{0.49\hsize}
	\includegraphics[width=\columnwidth]{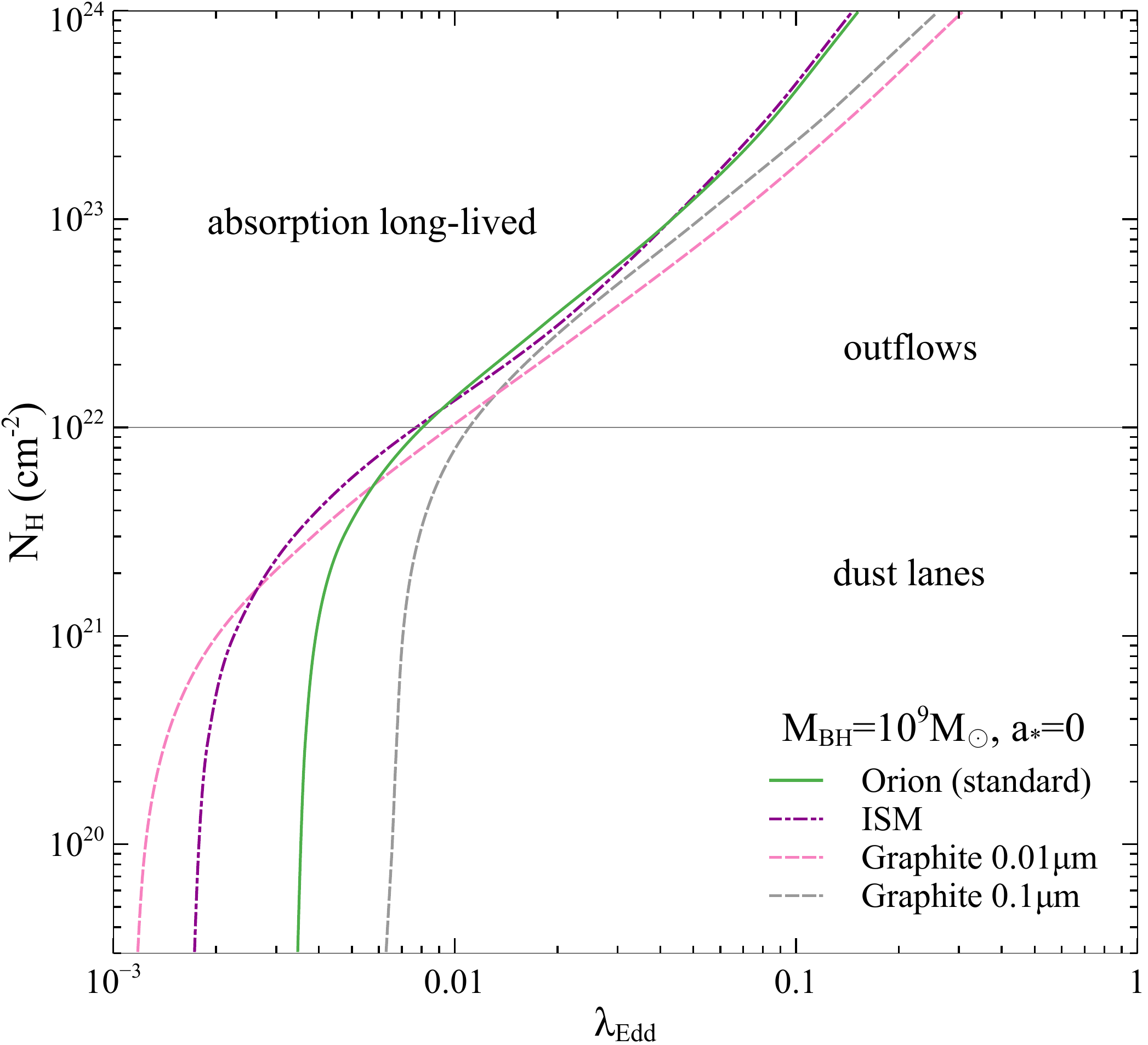}
\end{minipage}	
\begin{minipage}{0.49\hsize}
	\includegraphics[width=\columnwidth]{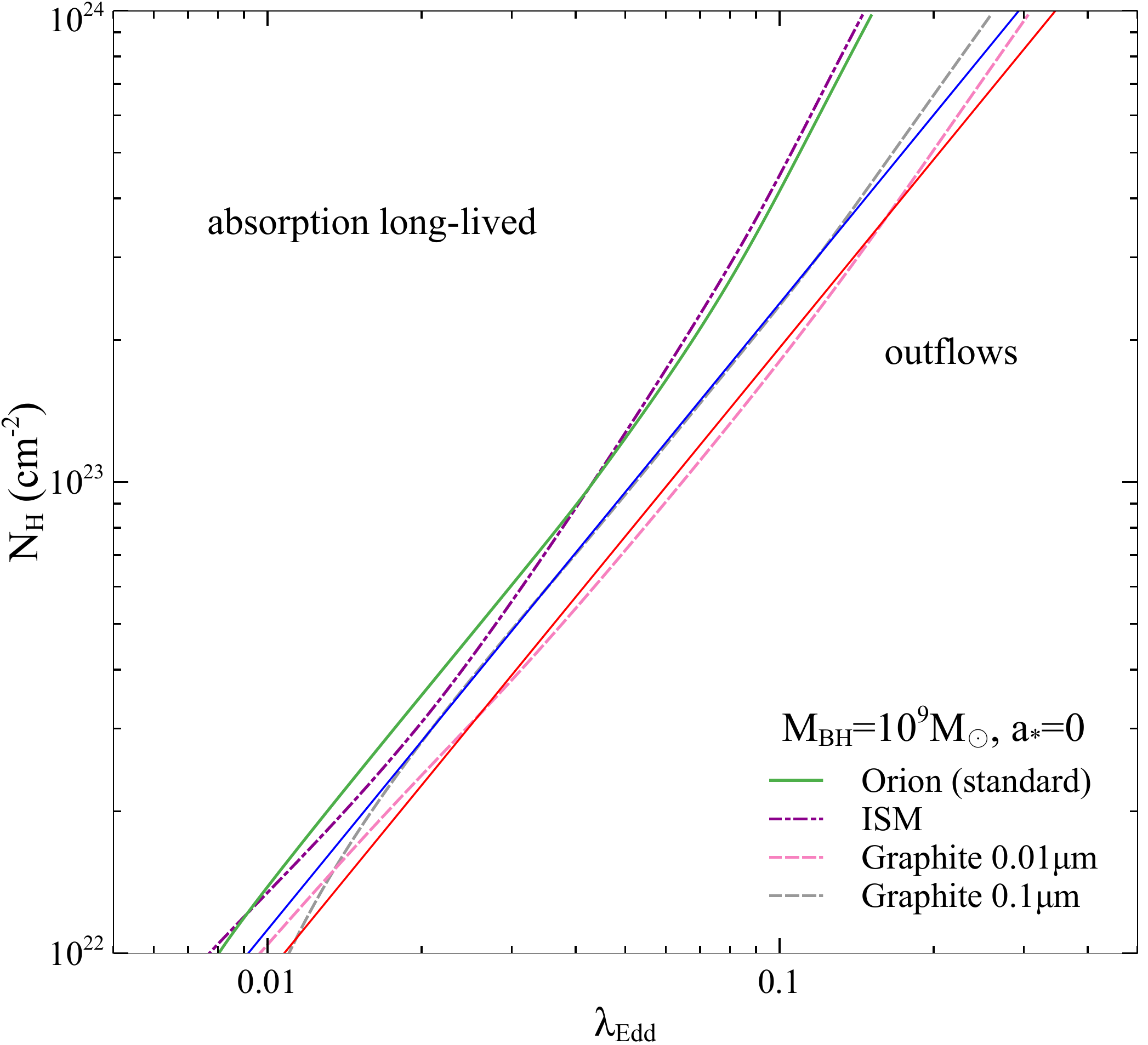}
\end{minipage}	
\vspace*{+0.5cm}

\caption{(Top left) The SEDs which explain how the incident AGN radiation is processed through the shell. No-grains model (cyan) has only ISM gas-phase abundance of $Z=0.5$ without grains. The SED coming out though the shell of Orion grain model (green) is shown as a sum of the transmitted radiation (brown) and diffused-out radiation (orange). 
(Top right) The spectra of four grain models of $Z=0.5$ and $N_{\rm H}=3\times10^{22}$ cm$^{-2}$ with the AGN of \mbh $=10^9$\msun and $a_* = 0$: 
Orion grain model (green); ISM grain model (dash-dot purple) applying the grains with a size distribution and abundance appropriate for the ISM of our galaxy which includes both a graphitic and silicate component; single-grain models of graphite, the size of which is $0.01\mu$m (dashed magenta) and $0.1\mu$m (dashed gray) respectively. See texts in Section \ref{sec:grain} for the detailed description of each model. Note that the dip near $0.006$ keV is seen for the graphite $0.01\mu$m model likely due to the $2175$ \AA\,extinction bump. (Bottom left) The \nhl diagram with the four grain models with $Z=0.5$: Orion model (green); ISM grain model (dash-dot purple); the others are single-grain models of graphite, the size of which is $0.01\mu$m (dashed magenta) and $0.1\mu$m (dashed gray). (Bottom right) Fitting power-law distributions to the two graphite models between the column densities of $10^{22}$ cm$^{-2}$ and $10^{24}$ cm$^{-2}$: $N_{\rm H}\propto\lambda_{\rm Edd}^{1.33}$ for the graphite $0.1\mu$m model (solid blue) and $N_{\rm H}\propto\lambda_{\rm Edd}^{1.30}$ for the graphite $0.01\mu$m model (solid red).}
 \label{fig:grain}
\end{figure*}

Principally dust grains absorb and scatter radiation which has a comparable wavelength to their size. In order to investigate the effects of grain size and abundance on the AGN spectrum and \nhl diagram, we tested out with four grain models: Orion grain model, ISM grain model and Single-grain models of graphite, the size of which is $0.01\mu$m and $0.1\mu$m, for $Z=0.5$ and $N_{\rm H}=3\times10^{22}$ cm$^{-2}$ with the AGN of \mbh $=10^9$\msun and $a_* = 0$. Their effects on the AGN spectrum and the diagram are shown in Fig.~\ref{fig:grain}.

The Orion grain model (green) indicates the graphitic and silicate grain abundance and its size distribution correspond to the values along the line of sight to the Trapezium stars in Orion with dust to gas ratio fitted to the Orion value. On the other hand, ISM grain model (dash-dot purple) applies the grains with a size distribution and abundance proper for the ISM of our galaxy which includes both a graphitic and silicate component. This is midway between the grey Orion grain opacities and the very blue Magellanic Clouds grains, as typical extragalactic grains. This model can reproduce the observed extinction properties with a ratio $R_{\nu}$ of extinction to reddening of $R_{\nu}=3.1$ for $Z=1$ while Orion model has $R_{\nu}=5.5$ for $Z=1$ which is almost like grey grains \citep{2006hbic.book.....F}. $R_{\nu}$ represents a measure of the size distribution, and large $R_{\nu}$ distributions like the Orion model lack small grains.

As shown in the top right of Fig.~\ref{fig:grain}, the Orion and ISM SEDs have a little difference while there would be differences in the unobservable FUV and XUV SEDs, which are caused by the grey cross-section dependence for the Orion grains, but those cannot be directly observable. In consequence, the Orion and ISM models show a negligible difference with respect to the outflow region in the \nhl diagram as shown in the bottom of Fig.~\ref{fig:grain}.

It is also shown that the UV and higher energy photons are significantly depleted by the grains due to absorption and scattering. The smaller the grain is, the more depletion of UV photons up to the lower energy can be seen. The large graphetic grain has smaller UV cross section about the same opacity in the red to IR. It should produce less acceleration as a result. ISM grains are much bluer, with more total opacity due to lots of low-mass small grains, so they would produce more acceleration. 

The acceleration produced by grains will be proportional to the grain cross section per unit mass of material, mostly hydrogen, since grains and gas are well coupled by Coulomb drag \cite{Spitzer.L78Physical-processes-in-the-interstellar-medium}.  Large-R grain mixtures have a larger fraction of small grains with larger cross section ($\propto \pi r^2$ where $r$ is a typical grain radius).  Smaller-R grains such as those found along lines of sight towards molecular clouds \citep{Bohlin.R81Ultraviolet-interstellar-extinction-toward} are dominated by larger radius grains and so have less opacity in the UV (see Figure 3 of \cite{Bohlin.R81Ultraviolet-interstellar-extinction-toward} ), so less radiative acceleration is produced. The effect of radiation pressure driving on dust-free ionized gas around an AGN has been considered by \citet{2002MNRAS.329..209M} and used to provide estimates of the black hole mass. The main agent is photoelectric absorption in a warm absorber.

\section{Summary and discussions}
\label{sec:discussion}

The results of our radiation simulations with various dusty gas and AGN properties, assuming a spherically-symmetric, constant-density, single dusty gas shell, reported in Section~\ref{sec:results} are the following:
\begin{enumerate}
	\item As expected, the higher the metal abundance, the more effective the radiative feedback is due to the increase of radiation pressure. If there is a NSC with a similar mass to the central black hole, the effective Eddington limit line shifts  to the right, due to less effective radiation pressure. (See Section~\ref{sec:abun}).
	\item The effects of the inner radius of the dusty gas shell, the shell width and the AGN bolometric luminosity on the \nhl diagram is negligible, less than 5\%. (See Section~\ref{sec:other}).
	\item The more massive the black hole is, the lower the energy goes to due to black-body radiation, while the higher the black hole spin is, the higher the energy shifts to in the AGN SEDs. Moreover, as \lam increases, the SED shifts to the higher energy. As the fraction of the energy at high energy, shifted to the UV from the optical band, increases, the coupling between radiation and dust could be increased.
	Another expected trend for all five \mbh models is that the more massive the black hole, the lower the UV fraction. (See Section~\ref{sec:sed}).
	\item The depletion of UV photons seen above \nh$\sim10^{22}$\cms reduces the acceleration of the outflowing material as the outer shell acts as dead weight. For a fixed \lam, the mean radiative acceleration of the shell is proportional to the black hole mass. Though spin dependence on that is negligible, if the UV photon fraction goes down as the spin increases, the radiative acceleration  drops while the peak energy increases. (See Section~\ref{sec:accel}).
	\item The higher the black hole mass , the smaller the dependence on spin, while the higher the spin, the larger dependence on the black hole mass seen in the \nhl plane. Overall, spin dependence on the effective Eddington limit line is relatively small. The reason why the slope in the \nhl plane for \lam $=0.01-0.1$ is getting steeper as \lam increases can be because the increase of the UV fraction makes the radiation pressure more effective. (See Section~\ref{sec:sed-nhl}).
	\item The smaller the grain is, the more depletion of UV photons up to the lower energy can be seen. The large graphetic grain has smaller UV cross section but about the same opacity in the red to IR. It should produce less acceleration. On the other hand, ISM grains are much bluer, with more total opacity due to lots of low-mass small grains, so they would produce more acceleration.  (See Section~\ref{sec:grain}).
\end{enumerate}

The major overall factors are the abundance of dust and the grain size distribution. We generally assume that the inner radius of the dusty shell is small. One sensible criterion, assuming no NSC, is that the dusty shell radius is within the region where the black hole mass is not exceeded by intervening stars. We can estimate that radius $r$ by assuming that the central regions of the galaxy are in an isothermal-sphere potential expressed as $M(r)=2\sigma^2 r/G$ and assuming a black hole mass -- stellar velocity dispersion ($M - \sigma$) relation. We find $r\sim 10 \sigma_{200}^{2.24}$~pc, using the relation of \cite{Gultekin2020}, where $\sigma_{200}$ is the stellar velocity dispersion in units of $200$ km/s.

 Some observational studies have discussed the location of outflowing sources in the \nhl diagram \citep[e.g.][]{2016AA...592A.148K, 2017Natur.549..488R, 2020MNRAS.495.2652L, 2020MNRAS.499.1823Z}, as well as theoretical studies \citep{2021MNRAS.502.3638I}. As discussed in Section \ref{sec:abun}, our new {\sc cloudy} simulation results are consistent with the analytic results of radiation trapping \citep{2018MNRAS.479.3335I}, as well as the previous IR-optically thin case \citep{2008MNRAS.385L..43F, 2009MNRAS.394L..89F}.

There are no recent data of sufficient quality to better reveal the forbidden region.  The new {\it Swift}/BAT data  \cite{2022arXiv220900014R}, confirm the  presence of the forbidden wedge, but large error bars leave its definition unclear. New relevant observational results should emerge from {\it eROSITA}, launched in July 2019. This all-Sky survey satellite with seven X-ray telescopes should uncover approximately 3 million AGNs \citep{2018SPIE10699E..5HP} when the eight complete sky scans are finished (4 are now complete but further progress is currently paused).


\section*{Acknowledgements}
We thank the referee for helpful comments. NA acknowledges the Gates Cambridge Scholarship from the Bill \& Melinda Gates Foundation. GJF acknowledges support by NSF (1816537, 1910687), NASA (ATP 17-ATP17-0141, 19-ATP19-0188), and STScI (HST-AR- 15018 and HST-GO-16196.003-A).

\section*{Data Availability}
No new data were generated or analysed in support of this research.



\bibliographystyle{mnras}
\bibliography{radiation} 




\appendix
\section{Additional figures}
\label{sec:appendix}

\begin{figure*}
\begin{minipage}{0.49\hsize}
	\includegraphics[width=\columnwidth]{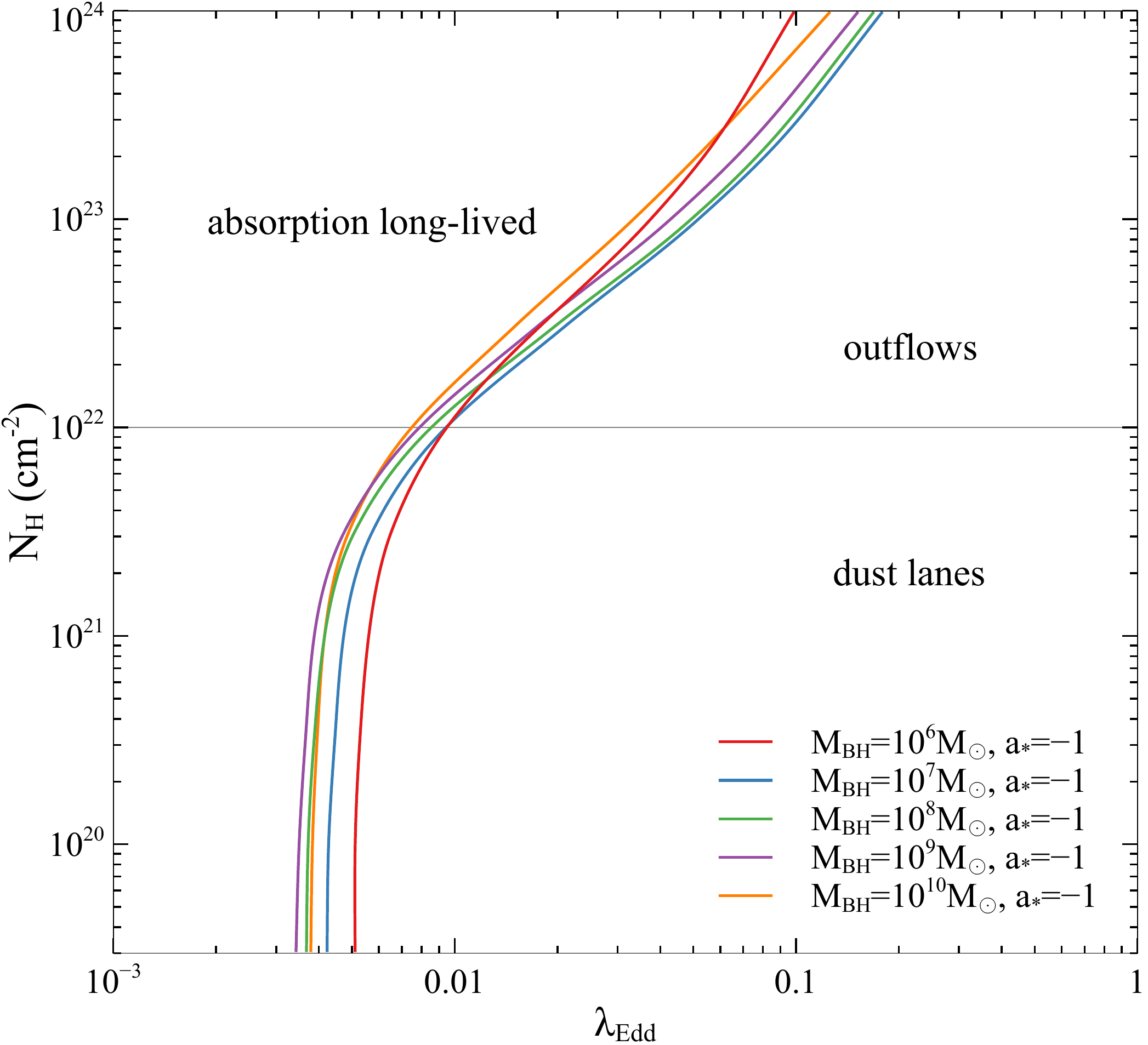}
\end{minipage}
\begin{minipage}{0.49\hsize}
	\includegraphics[width=\columnwidth]{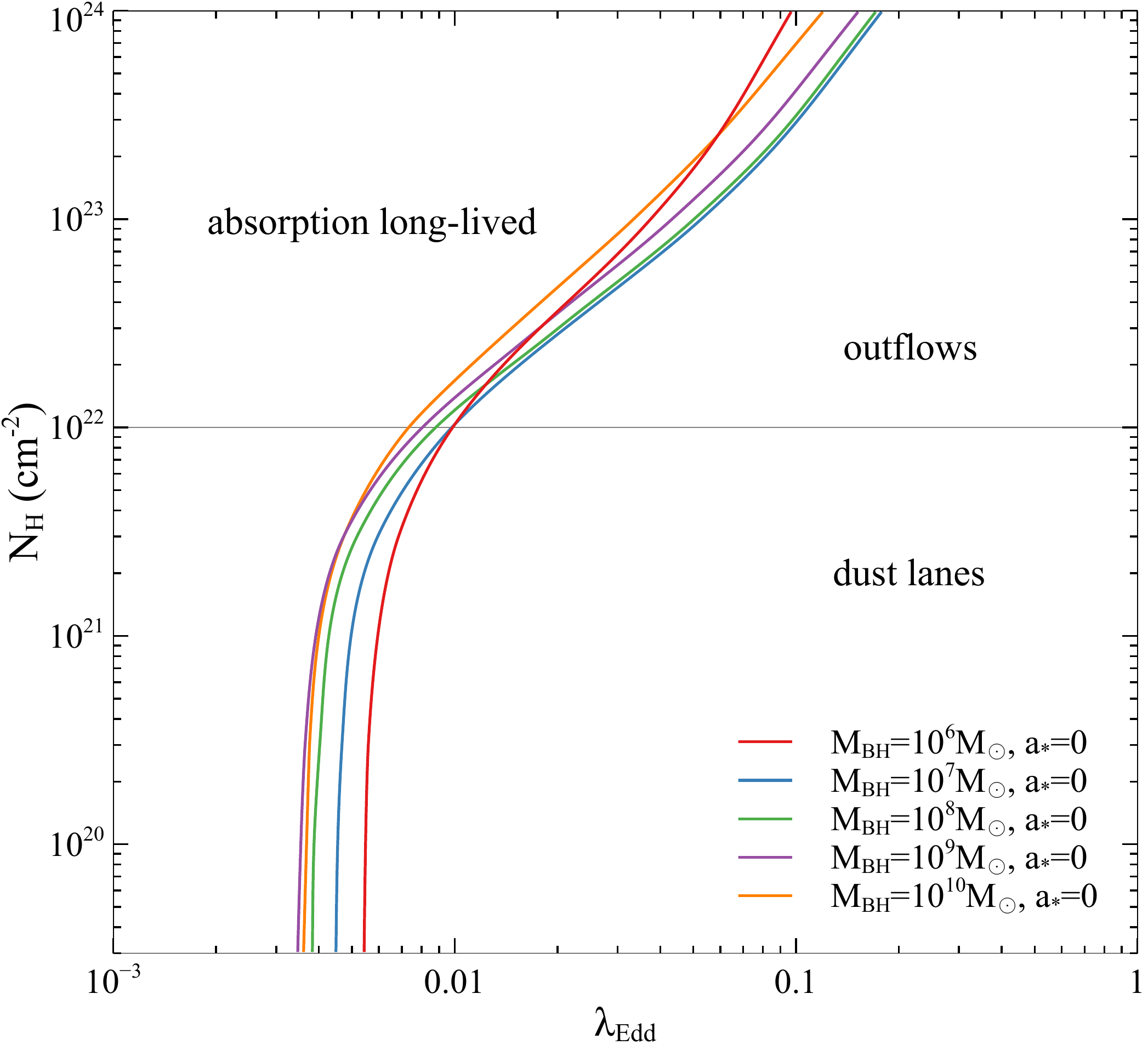}
\end{minipage}
\begin{minipage}{0.49\hsize}
	\includegraphics[width=\columnwidth]{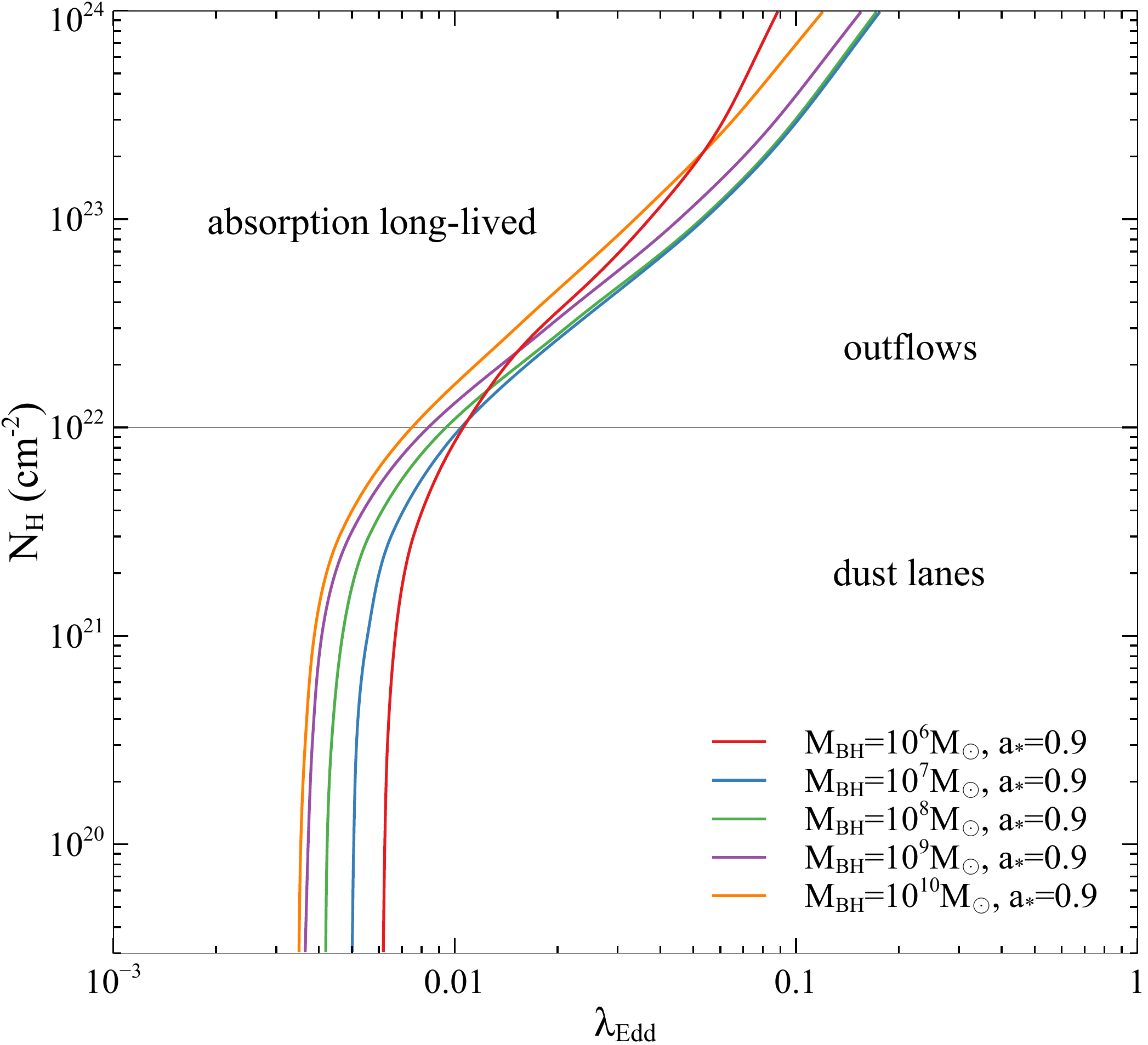}
\end{minipage}
\begin{minipage}{0.49\hsize}
	\includegraphics[width=\columnwidth]{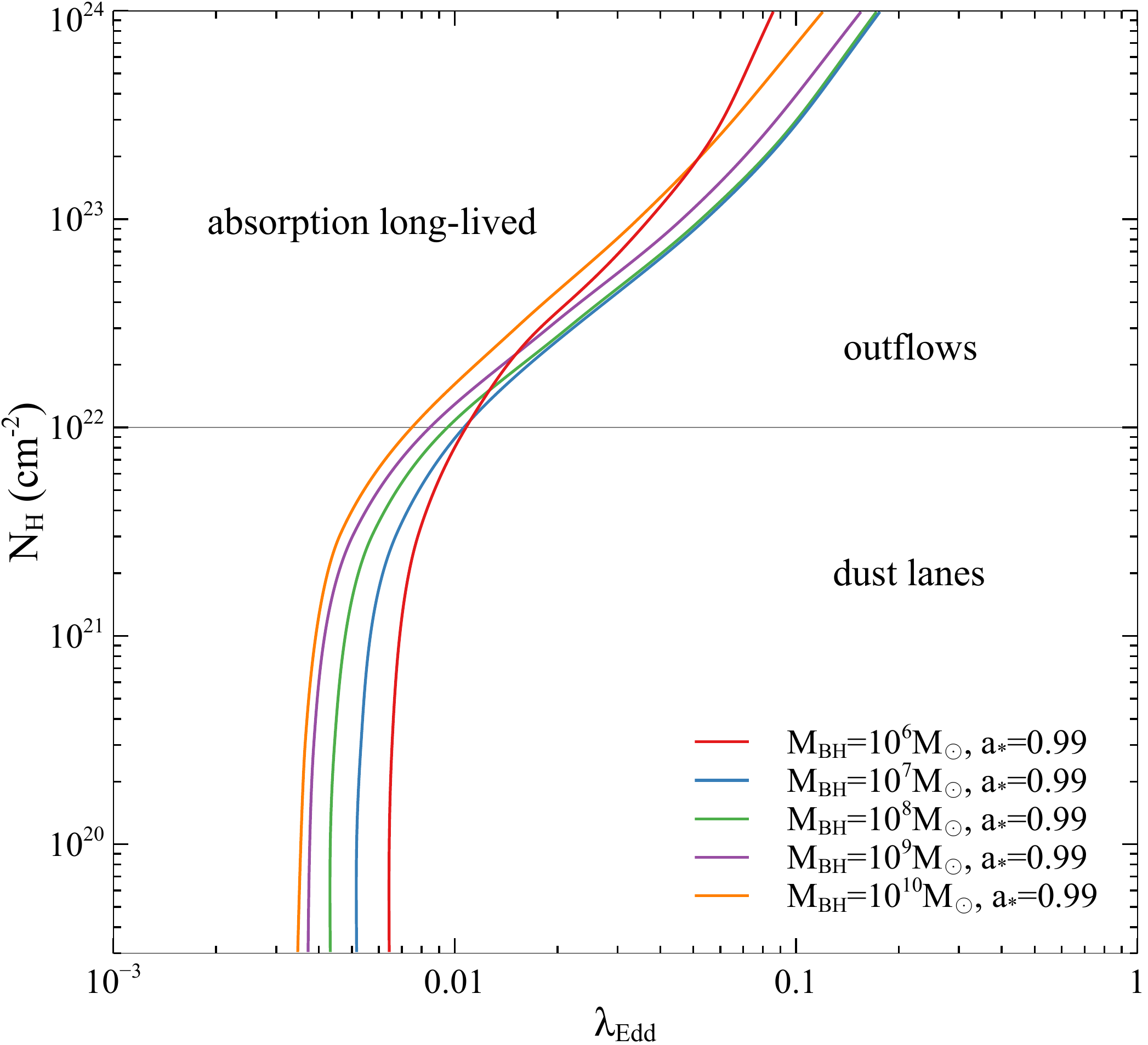}
\end{minipage}
\vspace*{+0.5cm} 
\caption{The \nhl diagram with spin parameters being fixed and different black hole masses: \mbh $=10^6$\msun (red), \mbh $=10^7$\msun (cyan), \mbh $=10^8$\msun (green), \mbh $=10^9$\msun (purple), \mbh $=10^{10}$\msun (orange), $a_* = -1$ (top left), $a_* = 0$ (top right), $a_* = 0.9$ (bottom left) and $a_* = 0.99$ (bottom right)}
 \label{fig:mbh}
\end{figure*}

\begin{figure*}
\begin{minipage}{0.47\hsize}
	\includegraphics[width=\columnwidth]{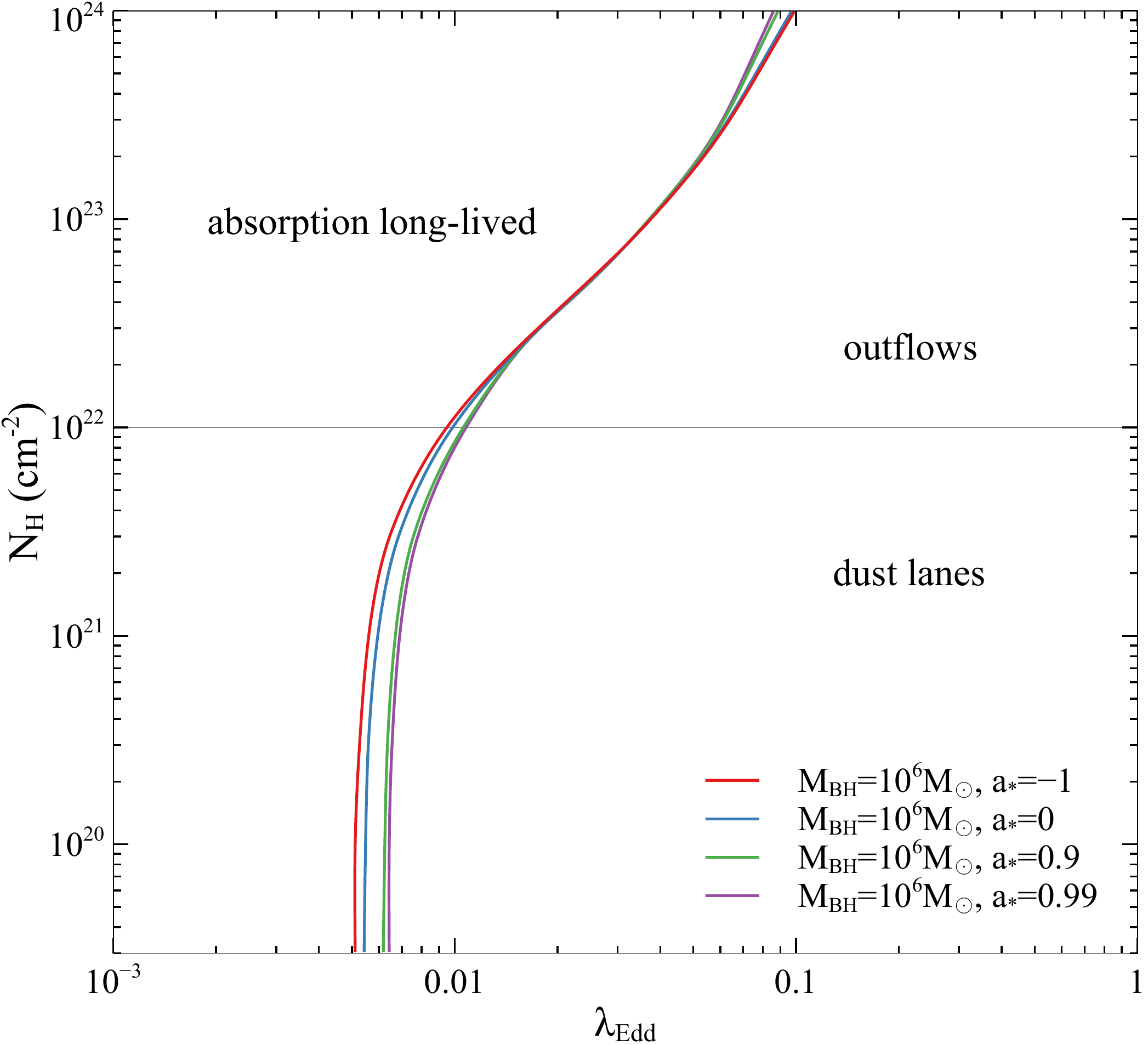}
\end{minipage}
\begin{minipage}{0.47\hsize}
	\includegraphics[width=\columnwidth]{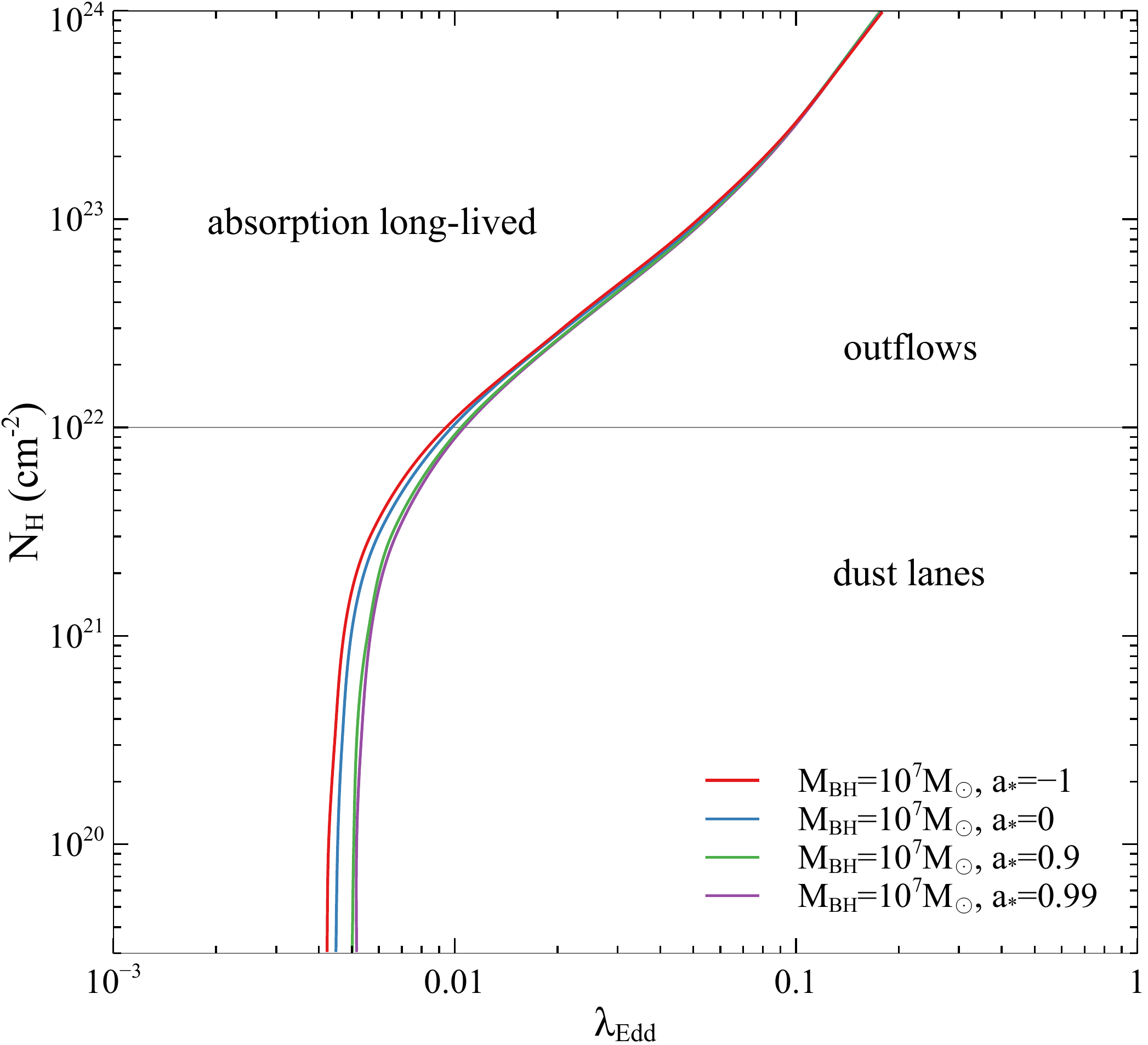}
\end{minipage}
\begin{minipage}{0.47\hsize}
	\includegraphics[width=\columnwidth]{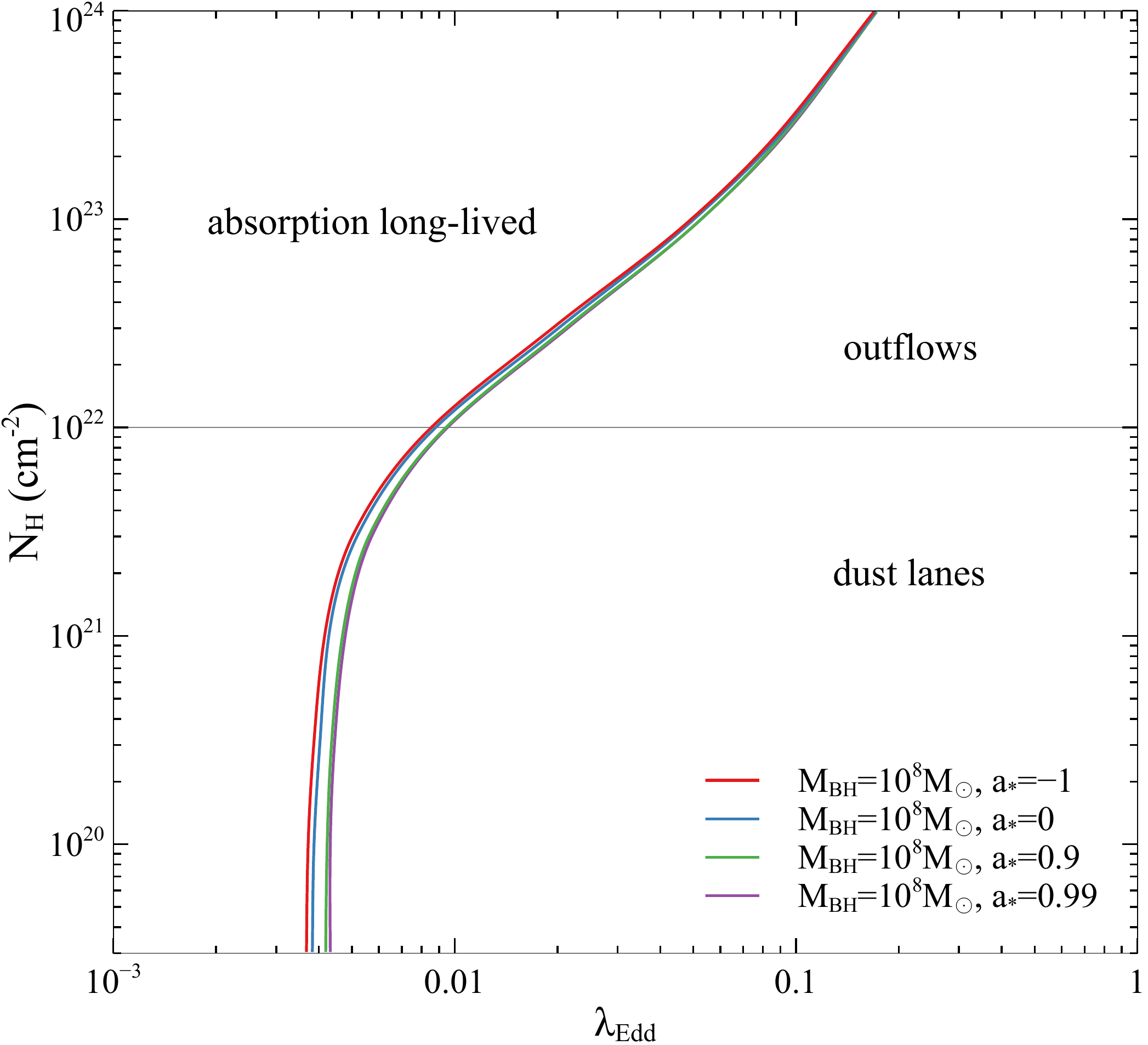}
\end{minipage}
\begin{minipage}{0.47\hsize}
	\includegraphics[width=\columnwidth]{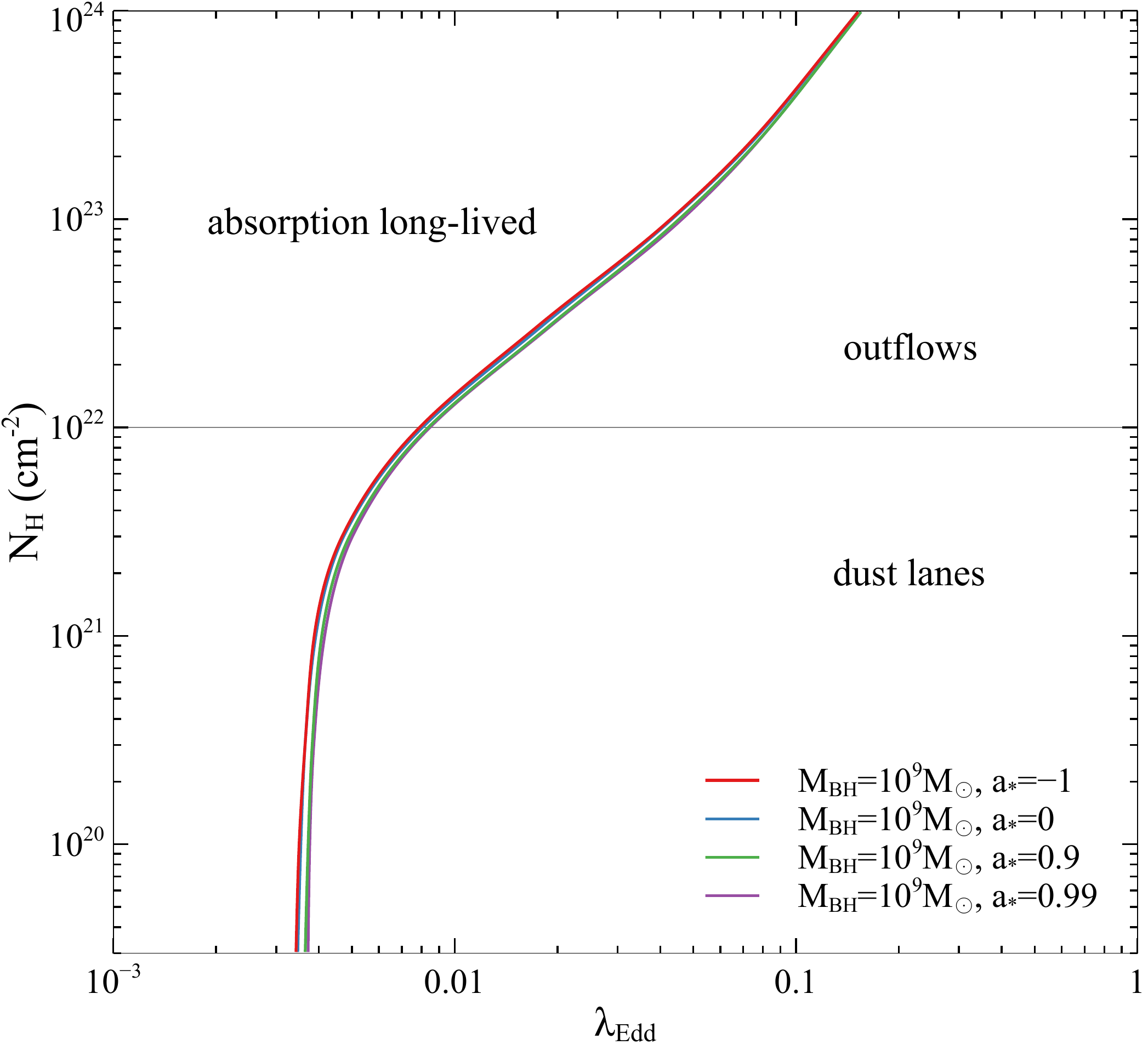}
\end{minipage}
\begin{minipage}{0.47\hsize}
	\includegraphics[width=\columnwidth]{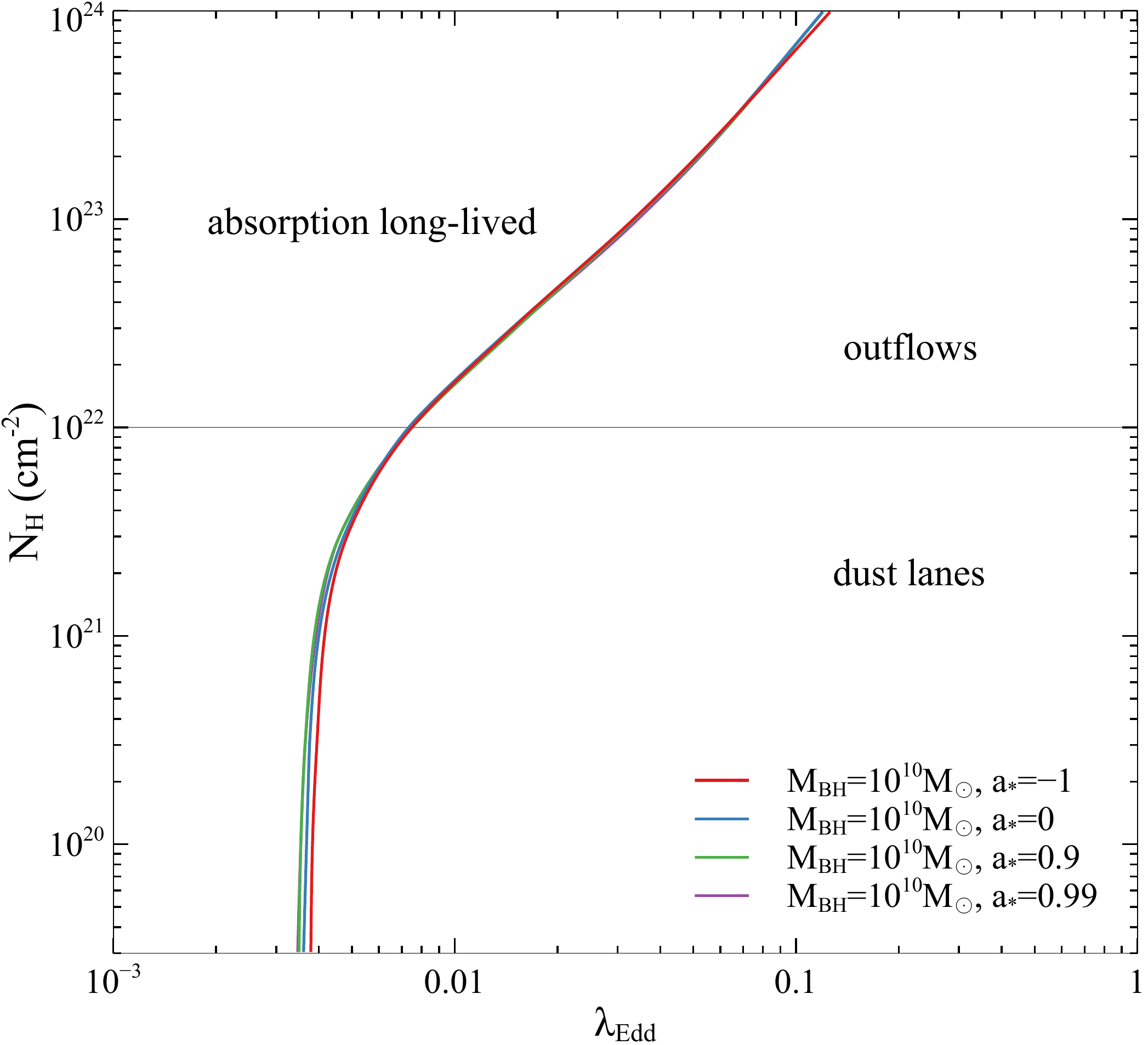}
\end{minipage}
\vspace*{+0.5cm} 
\caption{The \nhl diagram with fixed black hole masses and different spin parameters: $a_* = -1$ (red), $a_* = 0$ (cyan), $a_* = 0.9$ (green), $a_* = 0.99$ (purple), \mbh $=10^6$\msun (top left), \mbh $=10^7$\msun (top right), \mbh $=10^8$\msun (middle left), \mbh $=10^9$\msun (middle right) and \mbh $=10^{10}$\msun (bottom).}
 \label{fig:spin}
\end{figure*}


\bsp	
\label{lastpage}
\end{document}